\documentclass[fleqn,usenatbib]{mnras}

\usepackage{newtxtext,newtxmath}

\usepackage[T1]{fontenc}

\DeclareRobustCommand{\VAN}[3]{#2}
\let\VANthebibliography\thebibliography
\def\thebibliography{\DeclareRobustCommand{\VAN}[3]{##3}\VANthebibliography}

\usepackage{graphicx}	
\usepackage{amsmath}	
\usepackage{amssymb}	

\usepackage{caption}
\usepackage{subfig}
\usepackage[export]{adjustbox}
\usepackage{setspace}
\usepackage{multirow}
\usepackage{color}
\usepackage{ae,aecompl}
\usepackage{url}
\usepackage[para]{threeparttable}
\usepackage{gensymb}
\usepackage{textcomp}

\newsavebox{\imagebox}

\newcommand{\msun}{M$_{\odot}$}

\newcommand{\hi}{H{\sc i}}
\newcommand{\hii}{H{\sc ii}}
\newcommand{\Ha}{H$\alpha$}
\newcommand{\Hb}{H$\beta$}

\newcommand{\oiii}{[O{\sc iii}]}
\newcommand{\oii}{[O{\sc ii}]}
\newcommand{\nii}{[N{\sc ii}]}
\newcommand{\kms}{km s$^{-1}$}
\newcommand{\arc}{$^{\prime\prime}$}
\newcommand{\cgs}{erg s$^{-1}$ cm$^{-2}$}

\title[Inverted metallicity gradients in two Virgo star-forming dwarfs]
{Inverted metallicity gradients in two Virgo cluster star-forming dwarf galaxies: evidence of recent merging?\thanks{Based on observations collected 
at the Centro Astron\'omico Hispano Alem\'an (CAHA) at Calar Alto, operated jointly by the Max-Planck-Institut f\"{u}r Astronomie (MPIA) 
and the Instituto de Astrof\'isica de Andaluc\'ia (CSIC).}}
\author[M. Grossi et al.]{
M. Grossi$^{1}$\thanks{E-mail: grossi@astro.ufrj.br},
R. Garc\'ia-Benito$^{2}$,
A. Cortesi$^{1}$, D. R. Gon\c{c}alves$^{1}$, T. S. Gon\c{c}alves$^{1}$, P. A. A. Lopes$^{1}$,
\newauthor
K. Men\'endez-Delmestre$^{1}$, E. Telles$^{3}$\\
$^{1}$Observat\'orio do Valongo, Universidade Federal do Rio de Janeiro, Ladeira Pedro Ant\^onio 43, 20080-090 Rio de Janeiro, RJ, Brazil\\
$^{2}$ Instituto de Astrof\'sica de Andaluc\'ia (CSIC), PO Box 3004, E-18080 Granada, Spain\\
$^{3}$Observat\'orio Nacional, Rua General Jos\'e Cristino,77, 20921-400 Rio de Janeiro, RJ, Brazil \\
}

\date{}

\pubyear{2020}

\begin{document}
\label{firstpage}
\pagerange{\pageref{firstpage}--\pageref{lastpage}}
\maketitle

\begin{abstract}
We present integral field spectroscopy observations 
of two star-forming dwarf galaxies in the Virgo cluster (VCC~135 and VCC~324) obtained with PMAS/PPak at the Calar Alto 3.5 meter telescope. 
We derive metallicity maps using the N2 empirical calibrator. 
The galaxies show positive gas metallicity gradients, contrarily to what is usually found in other dwarfs or in spiral galaxies.
We measure gradient slopes of 0.20  $\pm$ 0.06 and 0.15  $\pm$ 0.03 dex/$R_e$ for VCC~135 and VCC~324, respectively.
Such a trend has been only observed in few, very isolated galaxies, or at higher redshifts ($z >$ 1). It is thought to be associated 
with accretion of metal-poor gas from the intergalactic medium, a mechanism that would be less likely to occur in a 
high-density environment like Virgo. We combine emission line observations with deep optical images to investigate the origin of the peculiar 
metallicity gradient. 
The presence of weak underlying substructures in both galaxies and the analysis of morphological diagnostics and of ionised gas kinematics
suggest that the inflow of metal-poor gas to the central regions of the dwarfs may be related to a recent merging event
with a gas-rich companion.
\end{abstract}

\begin{keywords}
galaxies: dwarf, galaxies: star formation, galaxies: ISM, galaxies: kinematics and dynamics, galaxies: interactions, galaxies: evolution.
\end{keywords}



\section{Introduction}
\label{sec:intro}

The abundance of heavy elements is a fundamental tracer 
of the evolutionary process of a galaxy.
The metal content depends on many factors such as star formation rate and gas mass fraction,
but it also reflects the interplay between outflows of chemically-enriched gas triggered by stellar feedback and external inflow
of gas from the intergalactic medium \citep[IGM;][]{2013ApJ...772..119L,2014MNRAS.443.3643P}.
\hii-region emission lines, tracing young and massive stellar populations, 
are linked to the most recent star formation events and they have been long used to 
investigate the gas-phase abundance pattern and evolution in galaxies 
\citep[][]{1971ApJ...168..327S,1979A&A....80..155L,1981ARA&A..19...77P,1998AJ....116.2805V,2004ApJ...613..898T,2010MNRAS.408.2115M}.

The advent of integral field unit (IFU) spectroscopic surveys 
such as CALIFA \citep[Calar Alto Legacy Integral Field Area;][]{2012A&A...538A...8S},
SAMI \citep[Sydney Australian Astronomical Observatory Multi-object Integral Field Spectrograph;][]{2012MNRAS.421..872C}, 
MaNGA  
\citep[Mapping Nearby Galaxies at Apache Point Observatory;][]{2015ApJ...798....7B}, and
AMUSING \citep[All-weather MUse Supernova Integral-field Nearby Galaxies;][]{2016MNRAS.455.4087G},
have made it possible to derive the spatial
distribution of metals for large samples of galaxies, using nebular lines.
Analysis of the radial distribution of heavy elements in galaxies is of particular interest because it allows us to understand  
the mass assembly history of a galaxy, the gas accretion process and its radial variation across the stellar disc, 
and the mechanisms that regulate the inside-out transportation of metals \citep{2001ApJ...554.1044C,2013MNRAS.435.2918M,2015MNRAS.448.2030H,2016A&A...587A..70S}. 

Studies of the gas metallicity radial gradient in local galaxies and of its dependence on stellar mass   
report different results.
The gradient slopes 
of a sample of 350 spiral galaxies from the CALIFA survey
show a slight dependence on M$_*$ and morphology,
with late-type low-mass objects (M$_* \sim 10^9 - 10^{9.5}$ \msun) displaying flatter trends \citep{2016A&A...587A..70S}.
Analysis of massive spirals (M$_* > 10^{10}$ \msun) in both the CALIFA and AMUSING surveys 
\citep{2014A&A...563A..49S,2018A&A...609A.119S} 
suggest that galaxies show a common abundance gradient characterised by an inner drop and an outer flattening 
with only a weak trend with stellar mass
\citep[for a more detailed review see][]{2020ARA&A..5812120S}.
From the resolved mass-metallicity relation, derived in a sample extracted from the MaNGA survey, 
\citet{2016MNRAS.463.2513B} infer that star-forming galaxies with stellar masses
above $10^{9.5}$ \msun\ display a common characteristic gradient.
On the other hand, using a MaNGA sample of 550 galaxies
\citet{2017MNRAS.469..151B} found that the gas metallicity gradients of star-forming systems  
 do depend on stellar mass:
the metal abundance decreases with radius at high M$_*$ -- the more massive the galaxies the steeper is the gradient --
whereas low-mass systems (M$_* \lesssim 10^{9.5}$ \msun) exhibit roughly constant metal content throughout their discs.

The environment where galaxies are evolving is another important factor defining the metal content and its radial distribution.
Low-mass satellites (M$_* < 10^{9.5}$ M$_{\odot}$) in higher density environments
tend to have higher gas abundances 
\citep{2014MNRAS.438..262P}. 
In very low-density environments there is mounting evidence of 
systems showing anomalous
metallicity gradients \citep{2014ApJ...783...45S}. 
Particularly, a class of extremely low-metallicity star-forming dwarfs 
with oxygen abundances 12 + log O/H $\lesssim$ 7.7 ($< 0.1$ Z$_{\odot}$),
present an off-center star-forming region of lower metallicity compared to the rest of the disc 
\citep{2011ApJ...743...77M,2016ApJ...819..110S,2017ApJ...835..159S,2018MNRAS.477..392L}. 
The decrease of the metal abundance in these regions is of the order of 0.3 dex or
larger and it is associated to the peak of the 
surface star formation rate \citep[SFR;][]{2014MNRAS.445.1104R,2015ApJ...810L..15S}.
Such metallicity drops 
may give evidence of pristine gas accretion from the IGM.

\begin{figure*}
\subfloat{
\begin{minipage}[t][][b]{.99\textwidth} 
  \includegraphics[bb=130 130 510 510,width=0.25\textwidth, clip,valign=t]{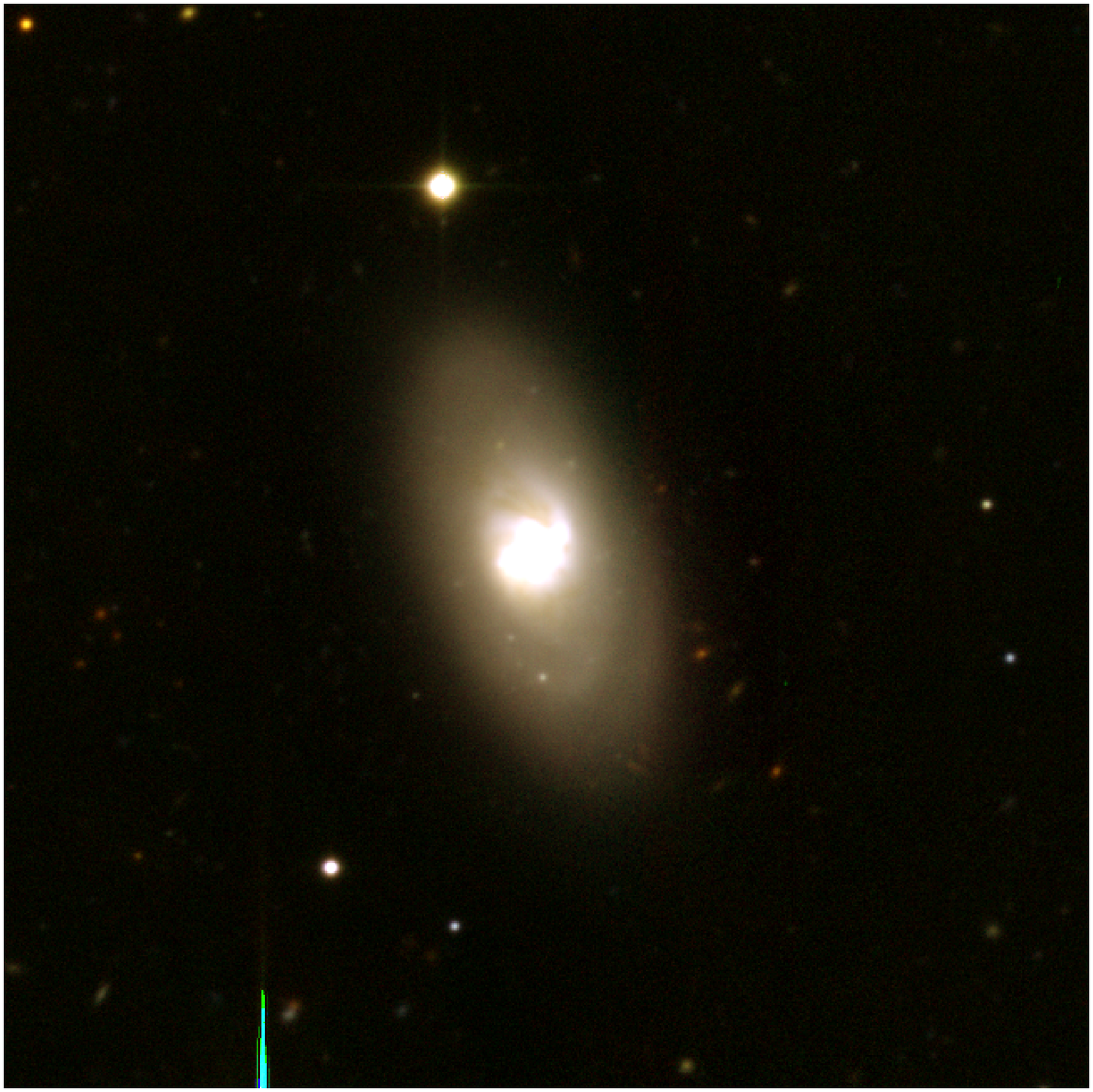}
  \includegraphics[bb= -30 -20 690 530,width=0.38\textwidth, clip,valign=t]{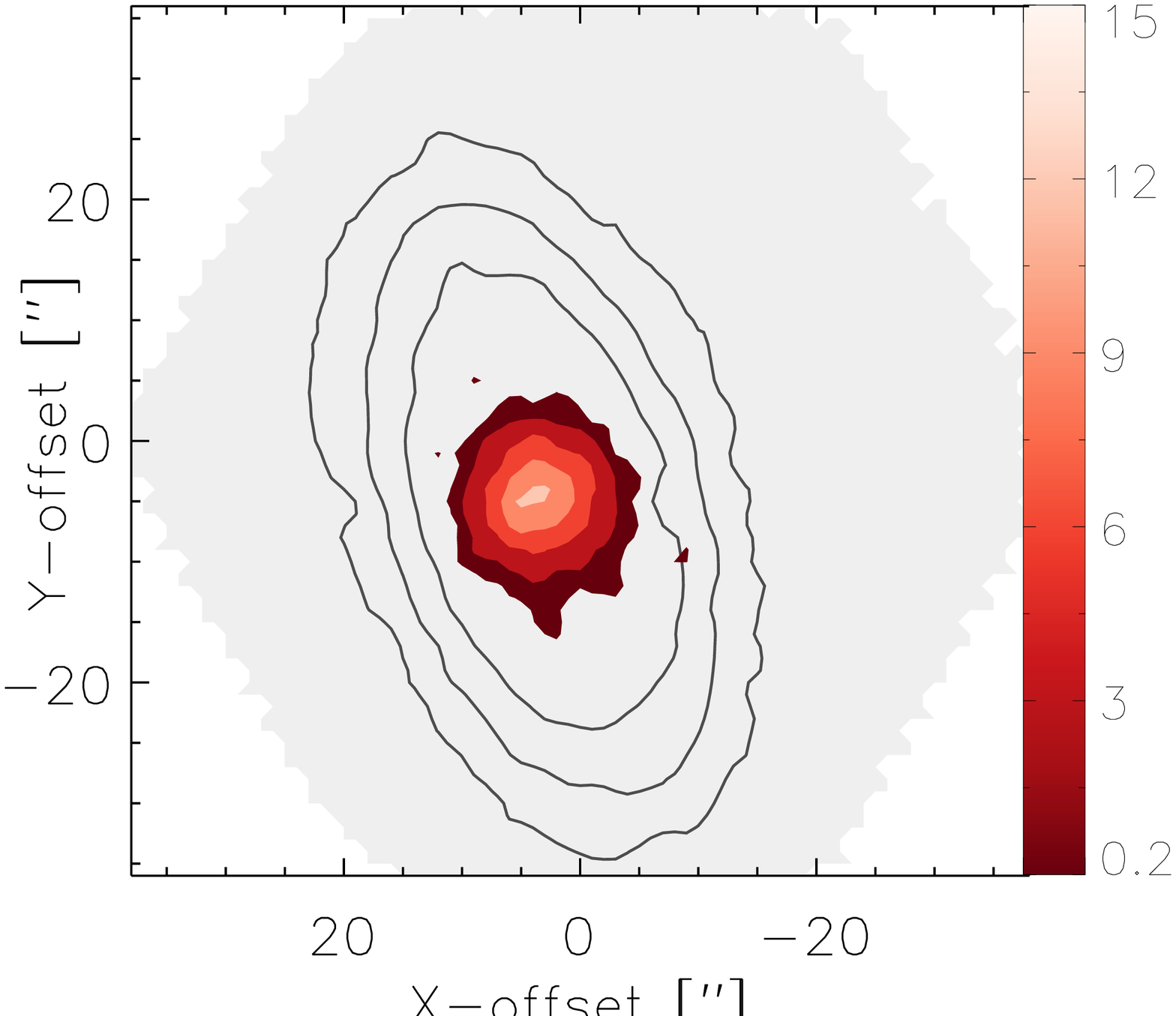}
  \includegraphics[bb= -30 -20 690 530,width=0.38\textwidth, clip,valign=t]{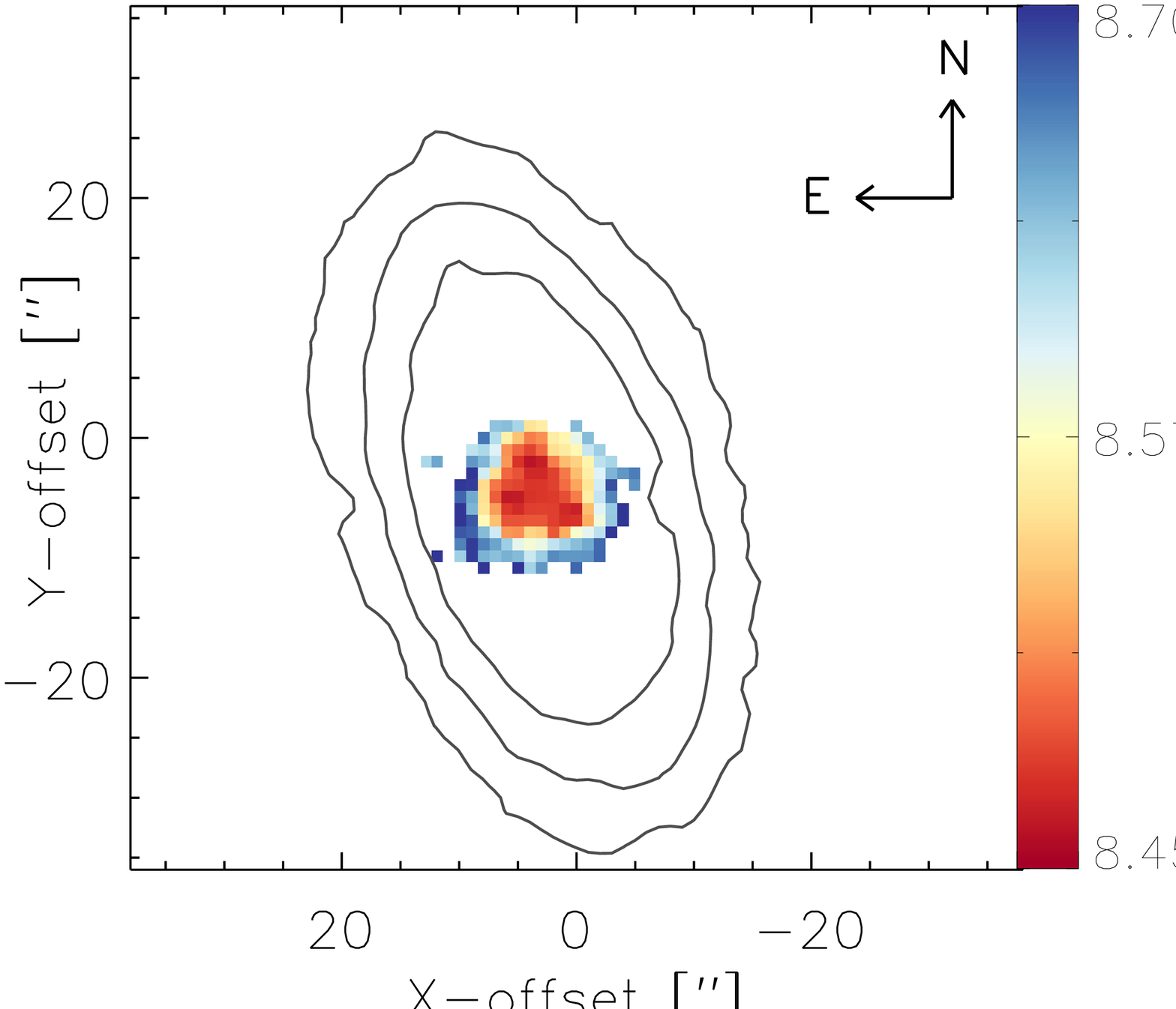} 
\end{minipage}}  \\
\subfloat{
\begin{minipage}[t][][b]{.99\textwidth} 
  \includegraphics[bb=140 100 520 470,width=0.255\textwidth, clip,valign=t]{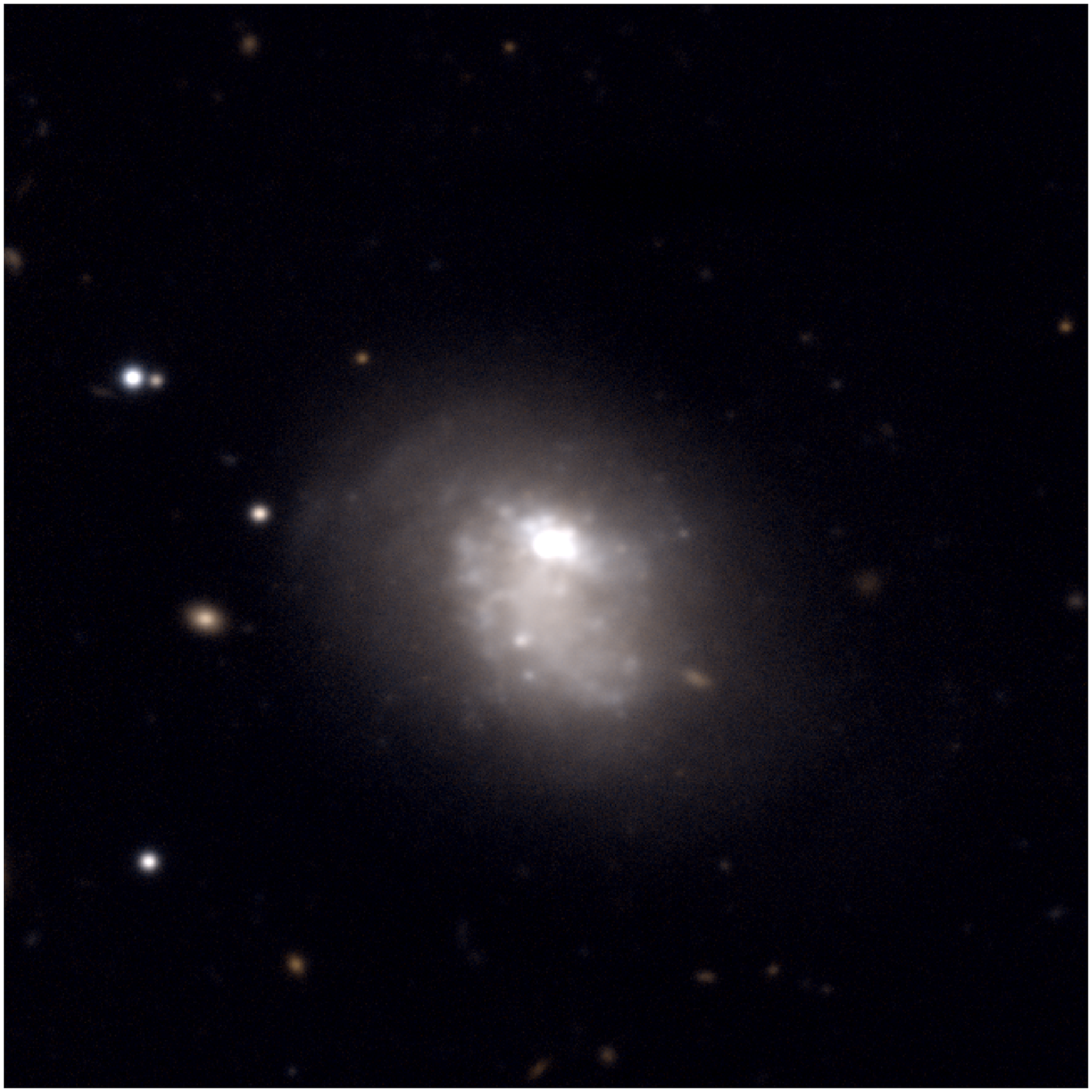}
  \includegraphics[bb= -30 -20 690 530,width=0.38\textwidth, clip,valign=t]{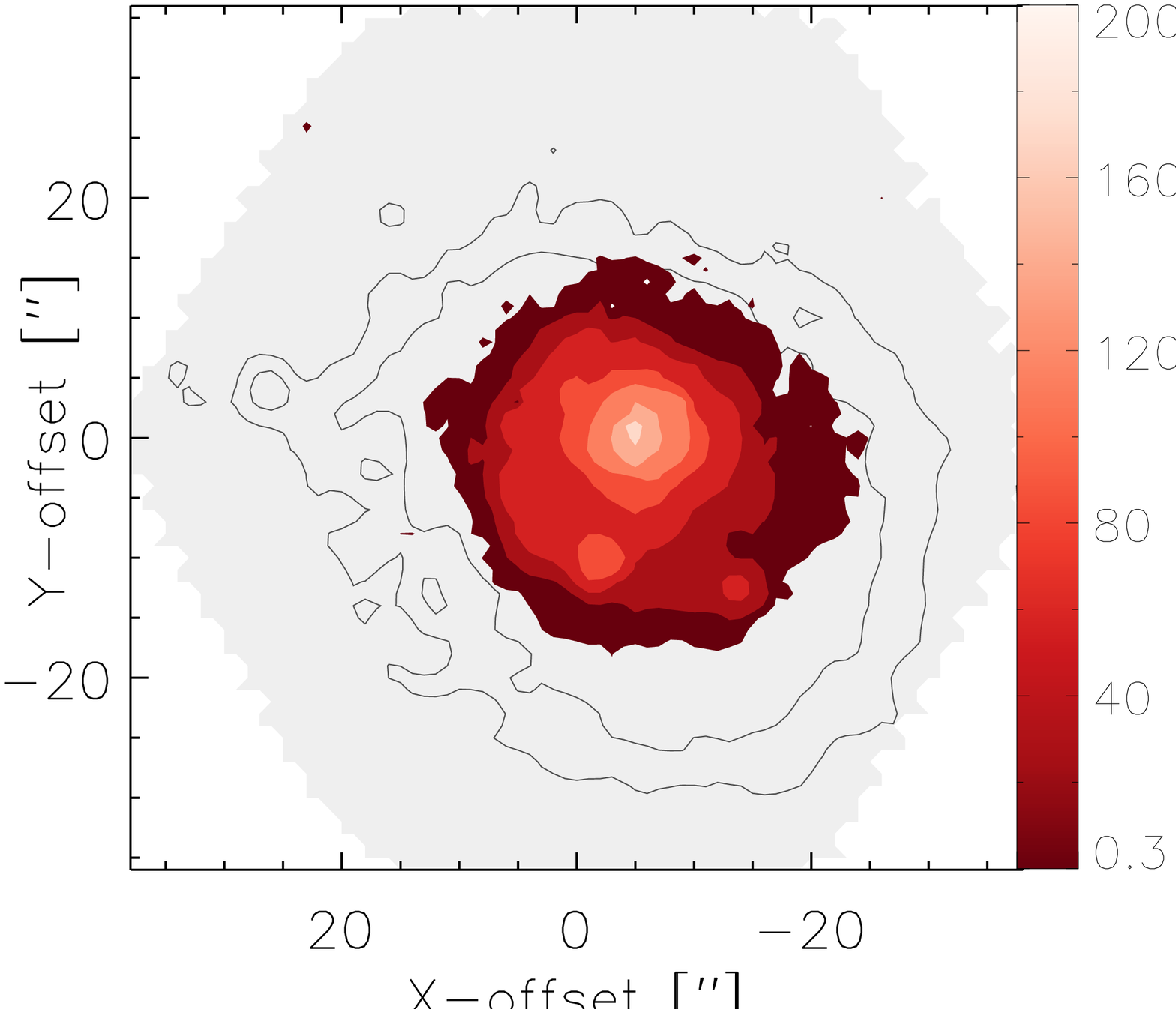}
  \includegraphics[bb=-30 -20 690 530,width=0.38\textwidth,clip,valign=t]{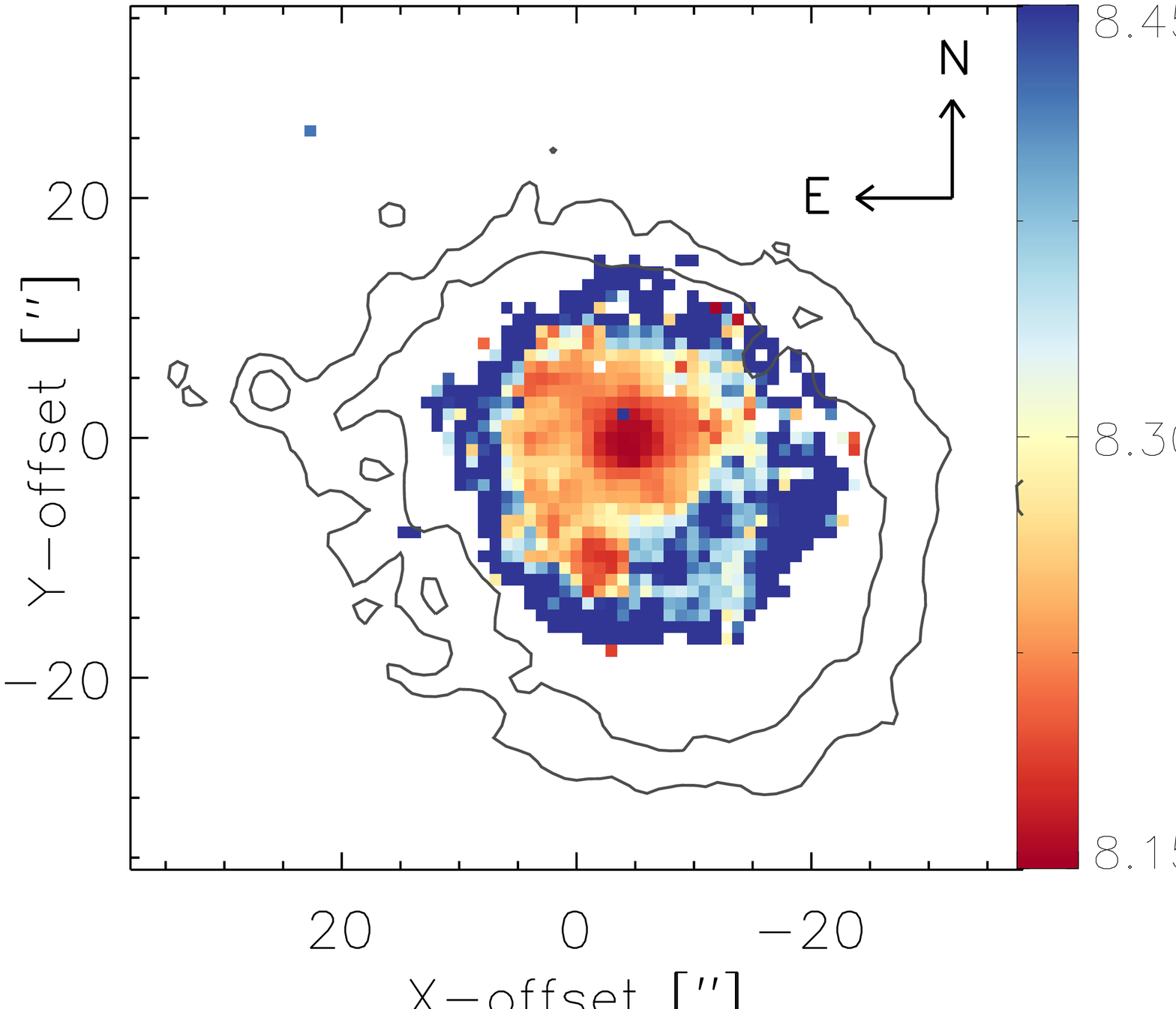}
\end{minipage}}  
       \caption{Left panels: $giz$ and $gr$ images of VCC~135 (top) and VCC~324 (bottom) 
    from the Next Generation Virgo Cluster Survey. The cutout size is 71\arc. 
    Central panels: H$\alpha$ emission of both galaxies overlaid on the contours of $g$-band continuum. Continuum contour 
    levels correspond to
    1,2,4 (VCC~135) and 1,2 (VCC~324) $\times$ 10$^{-18}$ \cgs\ arcsec$^{-2}$. The shaded region displays the PMAS/PPak FOV.
Right panels: Metallicity maps obtained with the N2 indicator \citep{2013A&A...559A.114M} for spaxels with EW(\Ha) 
above 3 \AA.}
\label{fig:met_img}
\end{figure*}

Positive or inverted metallicity gradients are observed at redshift $z \sim 1 - 2$
\citep[][]{2010Natur.467..811C,2012A&A...539A..93Q}.
They are
mostly attributed to the inflow of metal-poor gas from the IGM
diluting central metallicities. 
However, other scenarios require that gaseous outflows triggered by a central starburst
eject stellar nucleosynthesis products reducing the metal abundance \citep{2019ApJ...882...94W}.

Here we present the metallicity properties of two star-forming dwarf galaxies (SFDGs) in the Virgo cluster  
showing inverted oxygen abundance gradients: VCC~135 and VCC~324.
They are low-mass star-forming systems in a high density environment 
with different levels of \hi-deficiency \citep[i.e. from moderate to high,
][]{1984AJ.....89..758H,2013A&A...553A..90G}, implying that they are at different stages of interaction with the cluster.
To our knowledge these are the first examples of SFDGs with such characteristics in a cluster environment.
This is intriguing because accretion of cool gas 
from the IGM is expected to be suppressed
in dense environments, requiring other scenarios to be investigated.
The galaxies were selected from a sample of dwarfs detected by the Herschel Virgo Cluster Survey 
\citep{2013MNRAS.428.1880A,2015A&A...574A.126G} and
we obtained integral-field spectroscopy observations 
using the Potsdam Multi Aperture Spectrograph \citep[PMAS;][]{2005PASP..117..620R} in the PPak mode \citep{2006PASP..118..129K}
at the Calar Alto 3.5 meter telescope.
A large set of ancillary data
is available for these systems including observations of dust and molecular gas 
\citep{2015A&A...574A.126G,2016A&A...590A..27G}.

The paper is organised as follows: in Sect. \ref{sec:obs} we describe the observations and the data reduction procedure; 
in Sect. \ref{sec:gals} we present the properties of our target galaxies; the main results from the analysis of the IFU and 
optical data are 
given in Sect. \ref{sec:res}; Sect. \ref{sec:discuss} discusses possible
interpretations of the results: in Sect. \ref{sec:concl} we summarise our conclusions.

\begin{table*}
\begin{center}
\begin{threeparttable}
\begin{tabular}{lcccccccccc}
\hline \hline
ID      & $cz$   & $R_e$\tnote{\dag}  &  $i$         &     PA        & $\log$($M_*$)\tnote{\ddag} & $\log$($M_{\rm HI}$)\tnote{$\ddag$} & $\log$($M_{\rm H_2}$)\tnote{$\ddag$}  & $\log$(SFR)\tnote{$\ddag$} &  Def$_{\rm HI}$\tnote{$\ddag$} & $\nabla$log(O/H)  \\ 
        & [\kms] & [\arc]             & [$^{\circ}$] &  [$^{\circ}$] & [\msun]                    & [\msun]                             & [\msun]                               & [\msun\ yr$^ {-1}$]        &        [dex]                   & [dex/$R_e$]       \\ 
\hline \hline
VCC~135 & 2408  & 8.2   &  59 &  $20\pm2$ &  8.89 $\pm$ 0.04 & 6.64 $\pm$ 0.08 & 7.53 $\pm$ 0.08  & -1.04 $\pm$ 0.08 & 1.7 & 0.20  $\pm$ 0.06\\ 
VCC~324 & 1531  & 11.0  &  38 &  $51\pm2$ &  8.72 $\pm$ 0.04 & 8.23 $\pm$ 0.01 & 7.60 $\pm$ 0.26  & -0.75 $\pm$ 0.07 & 0.4 & 0.15  $\pm$ 0.03\\ 
\hline \hline
\end{tabular}
\begin{tablenotes}
\scriptsize
\item[$\dag$] \citet{2014ApJS..215...22K} \item[$\ddag$] \citet{2016A&A...590A..27G}
\end{tablenotes}
\end{threeparttable}
\caption{Properties of the selected Virgo star-forming dwarf galaxies assuming $d = 17$ Mpc.} 
\label{tab:prop}
\end{center}
\end{table*}

\section{Observations and data reduction}
\label{sec:obs} 

\subsection{PMAS/PPAK Observations}
\label{subsec:ifu} 

The PPak Integral Field Unit (IFU) consists of an array of 382 fibers arranged in a hexagonal field of view (FoV) of
$74^{\prime\prime} \times  65^{\prime\prime}$.
Observations were carried with both V500 and V1200 gratings, 
with a resolution of $\lambda/\Delta\lambda$ $\sim$ 850 at $\sim 5000$ \AA\ (FWHM $\sim 6$ \AA), 
and $\lambda/\Delta\lambda \sim$ 1650 at $ \lambda \sim$ 4500 \AA\ (FWHM $\sim$ 2.3 \AA), respectively.
Following the CALIFA survey observing strategy \citep{2012A&A...538A...8S}, the total integration time on each target was 1.5 h (V1200) and 0.75 h (V500).
A dithering scheme with three pointings was adopted to cover the complete FoV 
and to increase the
spatial resolution of the data. 
The two data sets were combined in one single cube (called ''COMBO``), covering the spectral range 3700 $-$ 7500 \AA\ with the
same resolution of the V500 grating \citep{2015A&A...576A.135G,2016A&A...594A..36S}. 
The spatial sampling is 1$^{\prime\prime}$, with a point spread function (PSF) given by a Moffat
function with a full width half maximum (FWHM) of 2\farcs5 and $\beta$ = 2.

Data reduction was performed using a python pipeline based on an upgraded version of 
\citet{2015A&A...576A.135G} and \citet{2016RMxAA..52...21S} and 
the data cubes were analysed with the spectral synthesis code \texttt{STARLIGHT} and the Python CALIFA
Starlight Synthesis organiser (\texttt{PyCASSO}) platform \citep{2013A&A...557A..86C,2017MNRAS.471.3727D}.
The spectra were fitted with the base of stellar libraries described in \citet{2017A&A...608A..27G}. 
In short, it combines the GRANADA
models of \citet{2005MNRAS.357..945G} for ages younger than 60 Myr
and the single stellar population (SSP) models from \citet{2015MNRAS.449.1177V} based on BaSTi isochrones for
older ages, giving a total of 254 SSPs. The $Z$ range covers eight
metallicities, log $Z$/$Z_{\odot}$ = $-$2.28, $-$1.79, $-$1.26, $-$0.66, $-$0.35, $-$0.06,
0.25, and +0.40, while the age is sampled in 37 bins per metallicity
varying from 1 Myr to 14 Gyr. The assumed initial mass function (IMF) is \citet{1955ApJ...121..161S}. Dust effects
were modeled using the \citet{1989ApJ...345..245C} reddening law with $R_V$ = 3.1.

The best stellar population model for each spectra was subtracted from the original cube 
to recover information on the distribution of the ionised gas. 
Line fluxes were measured with the \texttt{SHERPA} IFU line fitting software (\texttt{SHIFU}; Garc\'ia-Benito et al., in prep.), 
based on the package of \texttt{CIAO SHERPA} \citep{2001SPIE.4477...76F,2007ASPC..376..543D}. 
Maps of the \Ha, \Hb, \nii\
$\lambda$6584, \oiii\ $\lambda\lambda$4959,5007  emission and of the
line-of-sight velocity fields were generated.  
The COMBO cubes reach a S/N $\simeq$ 3 per spaxel and spectral pixel at 5600 \AA\ of $\sim$ 
2 $\times 10^{-18}$ erg s$^{-1}$ cm$^{-2}$ \AA$^{-1}$ 
arcsec$^{-2}$. 
The H$\alpha$ maps attain
a surface brightness limit of $\sim 10^{-17}$ erg s$^{-1}$ cm$^{-2}$ arcsec$^{-2}$ at a 3$\sigma$ level.

\subsection{MegaCam observations from the Next Generation Virgo Cluster Survey}
\label{subsec:megacam}

The Next Generation Virgo Cluster Survey (NGVCS)
is carried out with the MegaCam instrument on the the Canada--France--Hawaii Telescope
covering an area of 104 square degrees of the
Virgo cluster through 5 optical filters \citep[$ugriz$;][]{2012ApJS..200....4F}. The survey is designed to map the 
two main substructures surrounding the massive early-type galaxies M87 and M49  \citep[cluster A and B;][]{1987AJ.....94..251B},
out to their virial radii.
Details about the observation strategy and data reduction procedure can be
found in \citet{2012ApJS..200....4F}. 
With a 2$\sigma$ surface brightness limit of 
$\mu_g \sim$ 29 mag arcsec$^{-2}$ \citep{2016ApJ...824...10F} the NGVCS is the deepest optical survey of the cluster to date. 
The $g$-band
images are the most sensitive and with the least number of CCD artifacts compared to the other filters.
The MegaCam archive contains  
$ugiz$ and $gr$ observations of VCC~135 and VCC~324, respectively. 
The average seeing is 0\farcs7 (VCC~0135) and 0\farcs9 (VCC~324).
The pixel size of MegaCam images is 0\farcs187.

\section{Galaxy sample properties}
\label{sec:gals} 

\noindent {\bf VCC~135. } Classified as S$\,$pec/blue compact dwarf (BCD), it hosts a nuclear star-forming region 
overlaid on a red ($g - i \sim 1$ mag) stellar population \citep[][top-left panel of Fig. \ref{fig:met_img}]{2014A&A...562A..49M}. 
It is an extremely \hi-deficient system\footnote{The \hi\ deficiency parameter is defined as the logarithmic difference between 
the observed \hi\ mass and that expected for an isolated galaxy of the same morphological type:
Def$_{\rm HI} = \log M_{\rm HI}^{\rm ref} - \log M_{\rm HI}^{\rm obs}$ \citep{1983ApJ...267...35G,1984AJ.....89..758H}.} 
\citep[Def$_{\rm HI} = 1.7$;][]{2016A&A...590A..27G}, and
the \Ha\ emission
is compact, extending to only $\sim$ 1.5 times the effective radius ($R_ e =$ 8\farcs2; top-central panel of Fig. \ref{fig:met_img}).
The ionised gas is usually more centrally concentrated in \hi-deficient galaxies 
 \citep{2004ApJ...613..866K}.
VCC~135 has a large radial velocity ($cz$ = 2408 \kms)
compared to the cluster systemic velocity \citep[1149 \kms;][]{2018ApJ...865...40L}. Distances to
Virgo members are highly uncertain: according to \citet{2003A&A...400..451G} VCC~135 belongs to the M cloud, a substructure behind 
the cluster, while \citet{2014A&A...562A..49M} assume it is at the same distance as cluster A, $d \sim$ 17 Mpc. 
Using the infall model 
of \citet{1994ApJ...422...46P}, \citet{2014ApJS..215...22K} consider VCC~135 a possible Virgo member.
However its radial speed is about 200 \kms higher than the escape velocity predicted by the model at its
projected distance to the cluster centre. \\

\noindent {\bf VCC~324. }
The galaxy is classified as a BCD 
\citep{1985AJ.....90.1681B,2014A&A...562A..49M}.
The optical image shows a blue compact region superimposed on 
a lower surface brightness stellar component (Fig. \ref{fig:met_img}, bottom left).
The peak of the \Ha\ emission is located in correspondence of the compact 
region. Two bright knots are also visible to the south and south-west and appear to be connected by a tail
of ionised gas (bottom central panel of Fig. \ref{fig:met_img}). 
 \citet{2007ApJ...654..226R} suggest
that the galaxy is dominated by a recent star-forming event that started
less than 3.5 Myr ago. 
VCC~324 is moderately \hi\ deficient (Def$_{\rm HI} = 0.4$). 
\hi\ observations with the Very Large Array \citep[VLA;][]{1987ApJ...314...57L} show that the atomic gas extends out 
to about 2 times the optical radius ($\sim$ 1$^{\prime}$) and the
outer \hi\ distribution is elongated to the south-west.
The \hi\ velocity field is not well ordered and it shows only a weak hint
of rotation.
VCC~324 has a radial velocity $cz = 1531$ \kms and it is located in the Virgo southern extension, a substructure infalling towards the main cluster, 
at a projected distance of $\sim$ 5 degs from M49 ($\sim$ 1.5 Mpc at $d$ = 17 Mpc), the centre of cluster B \citep{1987AJ.....94..251B}. 
\citet{2014A&A...562A..49M} adopt a distance $d \sim$ 17 Mpc and according to the infall model of  \citet{1994ApJ...422...46P} 
VCC~324 is a possible Virgo member \citep{2014ApJS..215...22K}.

\section{Results}
\label{sec:res}

\subsection{Metallicity maps and gradients}
\label{subsec:metal}

The data products 
 provide spatially resolved information on the stellar population and 
the distribution of the ionised gas emission lines such as   
\Ha, \Hb, [O {\sc iii}]$\lambda$4363, [O {\sc iii}]$\lambda$5007, [N {\sc ii}]$\lambda\lambda$6548,6584, [S {\sc ii}]$\lambda\lambda$6717,6731. 
The [O {\sc iii}]$\lambda4363$ line is detected only in VCC~324 and the emission is concentrated in a very compact region 
compared to the size of the galaxy 
(see Fig. \ref{fig:app:oiii} and Appendix \ref{sec:app}). 
Therefore strong-line ratios provide the only tool
to derive oxygen abundances across the galaxy discs.
One of the most commonly used calibrations is based on the line ratio
N2 $=$ [N {\sc ii}]$\lambda6584$/\Ha\ \citep{2002MNRAS.330...69D,2004MNRAS.348L..59P}.
It has the advantage of being independent of reddening and its relation with O/H is single-valued although it 
saturates in the high-metallicity regime \citep{2006A&A...459...85N}. 
On the other hand, the N2-based calibration suffers from systematic uncertainties due to its dependence on the 
nitrogen-to-oxygen abundance (objects with high N/O ratios
will have higher O/H values than the real ones) and the degeneracy with the ionisation 
parameter  \citep{2009MNRAS.398..949P,2010IAUS..262...93S,2012MNRAS.426.2630L}. 
Another popular strong-line calibration is based on the line ratios 
O3N2 = ([O {\sc iii}]$\lambda5007$/H$\beta$)/([N {\sc ii}]$\lambda6584$/\Ha) \citep{2004MNRAS.348L..59P}.
However [O {\sc iii}]$\lambda5007$ emission of VCC~135 is even more compact than H$\alpha$, being detected to a 
radius $r \lesssim 0.6\,R_e$,
difficulting the use of O3N2 
to assess the metallicity spatial distribution in this galaxy.
Thus, despite the limitations of the N2 parameter, in the rest of this paper we discuss results based on this 
method that can be more easily applied to both our targets.
We adopt the calibration 
of \citet{2013A&A...559A.114M} to derive the oxygen abundances.
In Appendix \ref{sec:app} we discuss metallicity estimates of VCC~324 obtained with the O3N2 calibration and other diagnostics based on 
photoionisation models, and we show that our final conclusions are similar to what we infer with 
the N2 index. 

\begin{figure}
\begin{center}
\includegraphics[bb=-12 13 546 637,width=8.7cm, clip]{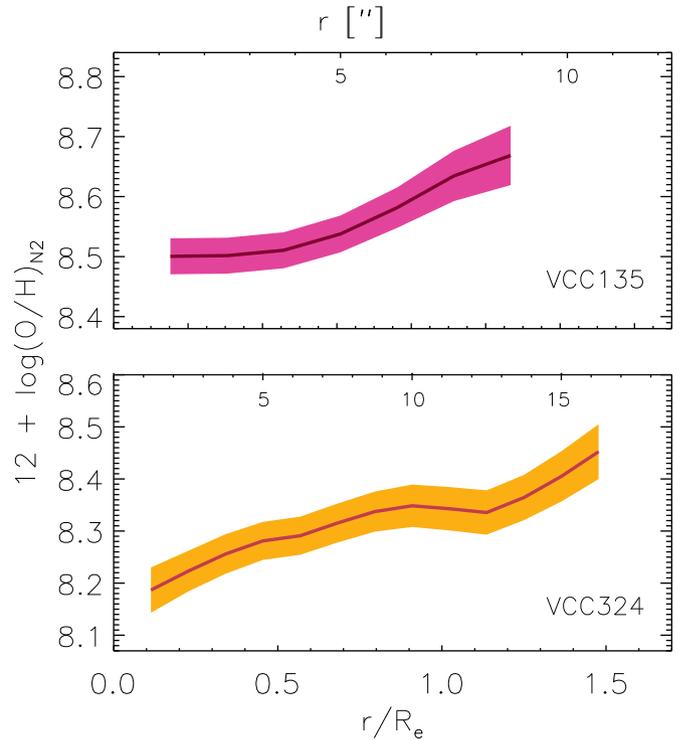}
\caption{Oxygen abundance radial profiles of  the two galaxies. In each panel the radial distance is shown in units of 
the disc effective radius $R_e$ in the bottom x axis and in arcseconds in the top x axis.
An inverted metallicity
gradient is detected in both galaxies.}
\label{fig:met_prof}
\end{center}
\end{figure}

The O/H maps of VCC~135 and VCC~324 are displayed in the right panels of Fig. \ref{fig:met_img}.
BPT diagnostic diagrams \citep{1981PASP...93....5B} for each spaxel where [O {\sc iii}]$\lambda5007$ was detected,
have been checked 
to ensure that the metallicity was obtained in regions ionised by  
star formation activity rather than by hot, old stars 
\citep{2016MNRAS.461.3111B}. 
Moreover, we required that the oxygen abundance was measured only in spaxels with \Ha\ equivalent width (EW) larger than 3\AA\ 
\citep{2011MNRAS.413.1687C,2018MNRAS.474.3727L}. This additional criterion excludes the outermost pixels 
where \Ha\ emission is detected in VCC~135 (Fig. \ref{fig:met_img}, top-central and top-right panels).
The maps show that the oxygen abundance 
of the regions with peak \Ha\ emission is lower by $\sim$ 0.2 dex compared to the outer part of the discs. 
An additional  
metallicity dip can be seen in VCC~324 at about 10\arc\ to the south with 12 + log(O/H) $\sim$ 8.25 dex,
in correspondence of a local enhancement of the \Ha\ emission.
An anticorrelation between the peak of the star formation rate and the oxygen abundance 
is usually observed in
galaxies with metallicity drops in their discs 
\citep{2010Natur.467..811C,2014MNRAS.445.1104R,2016MNRAS.456.1549L,2018MNRAS.476.4765S,2020MNRAS.492.6027E}
 and we discuss in Sect. \ref{sec:discuss} what are the possible scenarios to explain this trend.

To derive the metallicity gradients, we used the \texttt{ELLIPSE} task within \texttt{IRAF}. We defined a series of 
elliptical apertures on the \Ha\ and [N{ \sc ii}] maps with increasing radii taken
in steps of 1\farcs25, so that
the sampling is done at separations corresponding to half width of the PSF\footnote{ 
Ideally the width of the annuli should be comparable to the spatial resolution of roughly 2\farcs5,
however we chose this value to obtain a reasonably sampled profile of the abundance of VCC~135 where the ionised
gas emitting region is very compact.
}. Ellipsis axial ratios and position angles (PAs)
were inferred from the analysis of the SDSS 
$r$-band images with \texttt{ELLIPSE} and they are displayed in Tab. \ref{tab:prop}.
The apertures 
were centred on the peak of the \Ha\ emission.
The uncertainty on the metallicity within each
bin takes into account the average errors on the emission-line fluxes along the best-fit isophote as well as the errors on 
the calibration coefficients \citep{2013A&A...559A.114M}. 

Figure \ref{fig:met_prof} shows  the metallicity variation as a function of radial distance. Both galaxies 
display positive gradients with O/H increasing from the main star-forming region
outwards.
A linear fit provides slopes of 0.20 (0.27) $\pm$ 0.06 and  
0.15 (0.16) $\pm$ 0.03 dex/$R_e$ (dex/kpc) for VCC~135 and VCC~324, respectively. We assumed $d$ = 17 Mpc, 
$R_e^{\rm VCC135} =$ 8\farcs2, and $R_e^{\rm VCC324} = $ 11\arc\ in the $r$ band \citep[Table \ref{tab:prop};][]{2014ApJS..215...22K}. 
As a comparison, dwarf galaxies at $z \sim 2$ show flatter positive radial 
gradients, with $\nabla$log(O/H) $\sim 0.11 \pm 0.01$ dex/kpc \citep{2019ApJ...882...94W}. Inverted 
gradients are also detected in a 
subsample of 20 spiral galaxies in the CALIFA survey with an average flatter slope than our dwarfs, $\nabla$log(O/H)
 = 0.048 $\pm$ 0.033 dex/$R_e$ \citep{2016A&A...595A..62P}.


\subsection{Ionised-gas velocity curves}
\label{subsec:vel}

PMAS/PPak observations allow us to extract two-dimensional velocity maps of both the gas and stellar components. 
Studying the ionised gas 
kinematics  can help inferring evidence of ongoing interaction processes traced by
velocity field perturbations or anomalies \citep{2014MNRAS.439..284R,2015A&A...582A..21B}.

\begin{figure}
\begin{flushleft}
\subfloat{\includegraphics[bb= 35 15 580 260,width=7.7cm, clip]{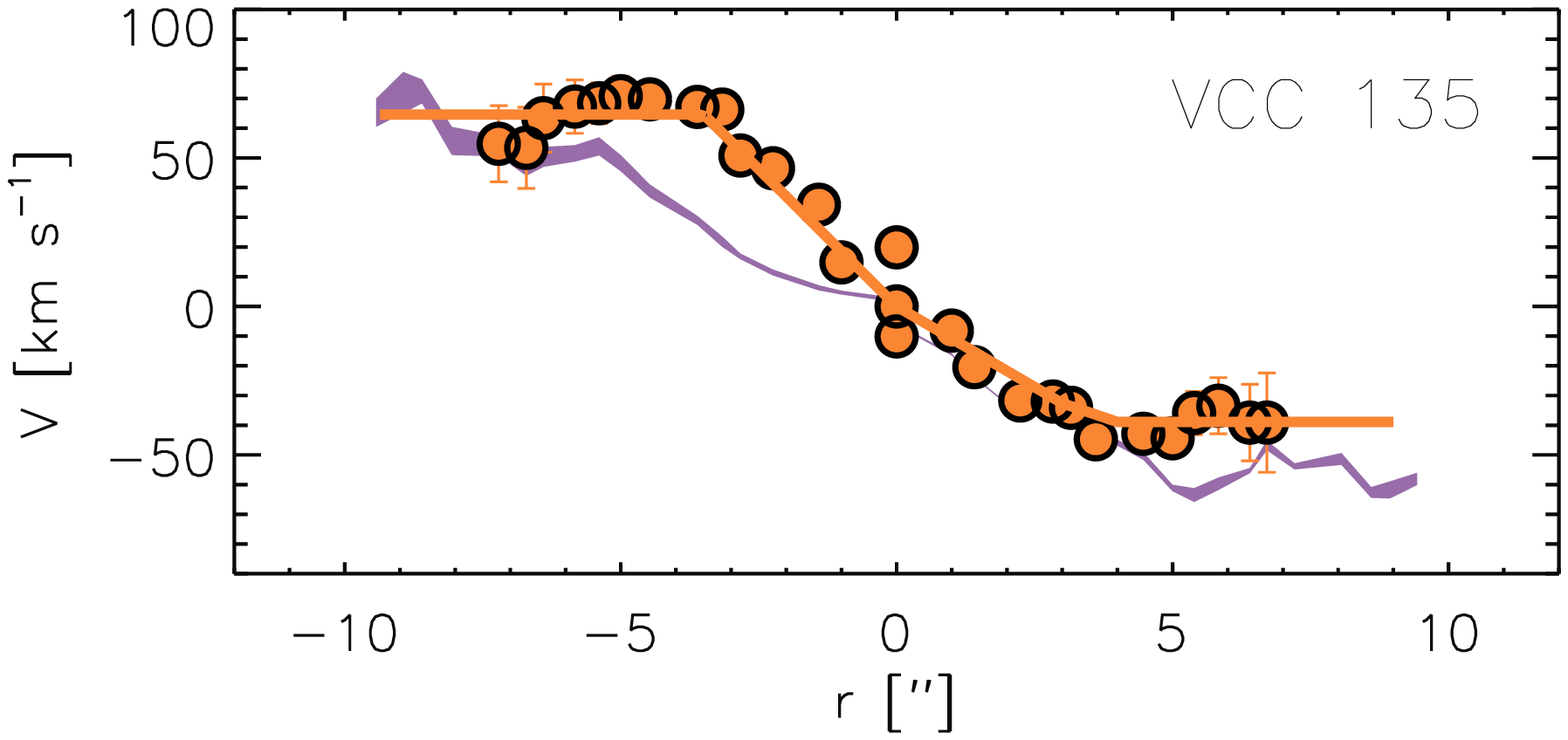}}  \\
\end{flushleft}
\centering
\subfloat{\includegraphics[bb= 10 -15 650 520,width=8.4cm, clip]{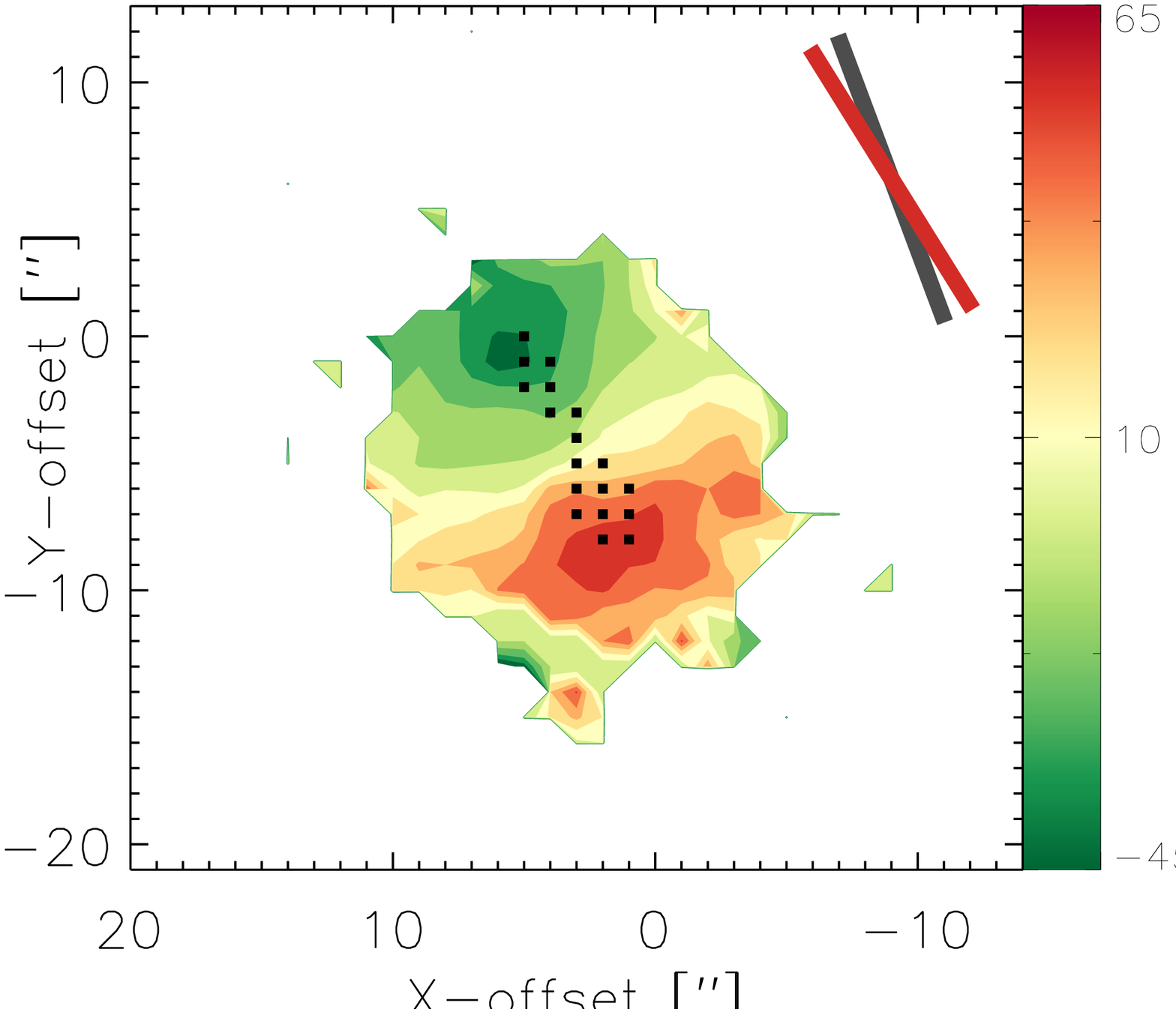}}  
\caption{Bottom: \Ha\ velocity map of VCC~135. Contours correspond to a step size of 15 \kms. 
Black squares indicate the spaxels used to determine the PA of the kinematic axis.
The two bars at the top-right corners compare 
the position angles of the photometric (grey) and kinematic axis (red). 
Top: Position-velocity diagram of the ionised gas of VCC~135 for spaxels in an 
artificial slit of 2$^{\prime\prime}$ width centred on the KC with PA$_{\rm kin}$ = 211$^{\circ}$ (filled circles).
The solid line shows the best-fit rotation curve model (see text). 
The filled (purple) curve indicates the position velocity diagram of the stellar component estimated in the same spaxels.
}
\label{fig:vel135}
\end{figure}

\begin{figure}
\begin{flushleft}
\subfloat{\includegraphics[bb= 35 15 580 260,width=7.7cm, clip]{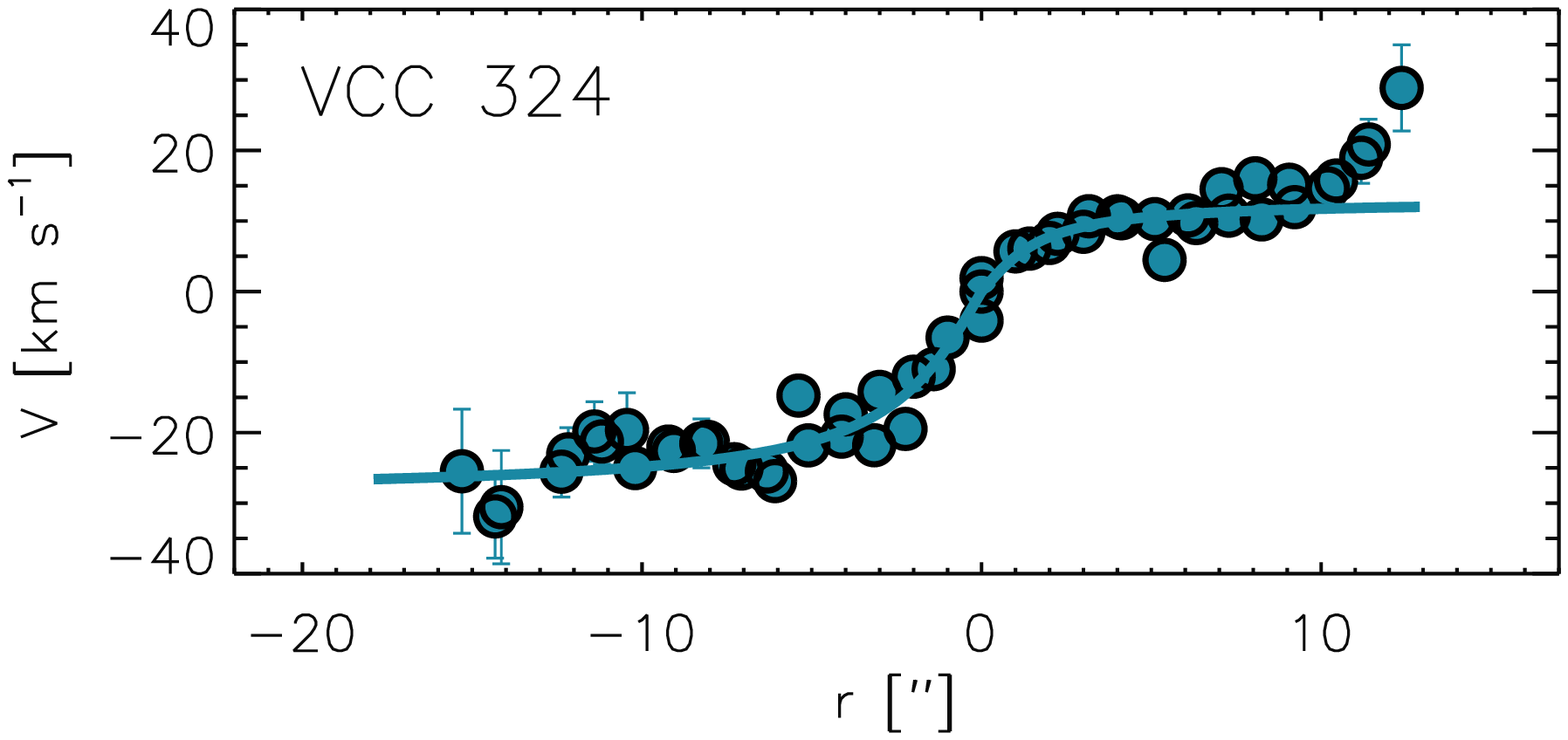}}  \\
\end{flushleft}
\centering
\subfloat{\includegraphics[bb= 10 -15 650 520,width=8.4cm, clip]{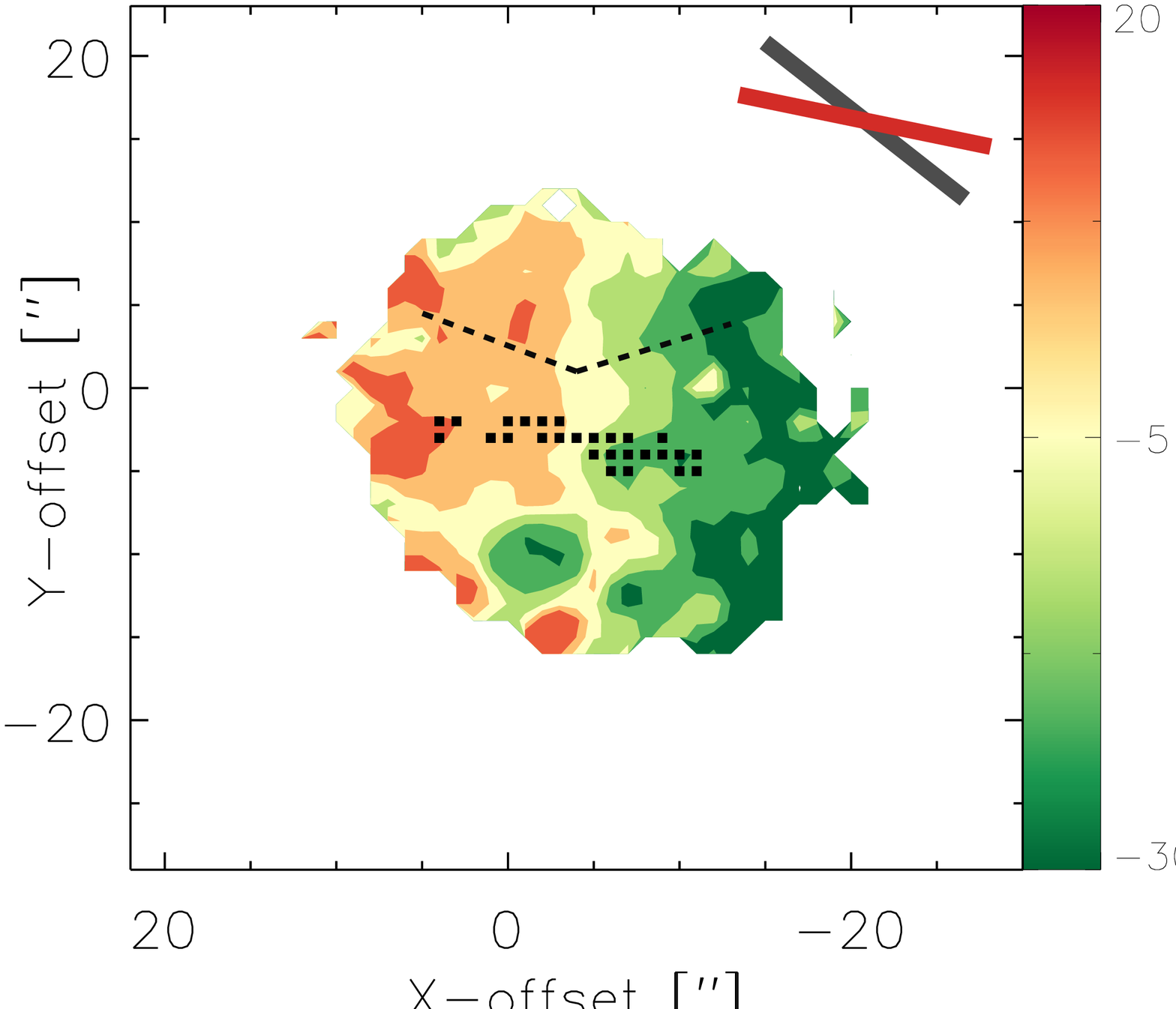}}  
\caption{Bottom: \Ha\ velocity map of VCC~324. Contours correspond to a step size of 10 \kms.
Black squares indicate the spaxels used to determine the PA of the kinematic axis, assuming as KC the centre
of symmetry of the gas distribution (KC1).
The dashed lines show the estimated receding and approaching kinematic axis assuming that the KC corresponds to 
the peak of the \Ha\ emission (KC2).
The two bars at the top-right corner indicate the position angles of the photometric (grey) and kinematic axis (red). 
Top: Position-velocity diagram of the ionised gas of VCC~324 for spaxels in an 
artificial slit of 2$^{\prime\prime}$, centred on KC1, with PA$_{\rm kin}$ = 78$^{\circ}$ (filled circles).
The solid line shows the best-fit rotation curve model (see text).}
\label{fig:vel324}
\end{figure}

The ionised-gas velocity maps indicate that both galaxies display a velocity gradient 
(Figs. \ref{fig:vel135} and \ref{fig:vel324}).
To derive the ionised gas rotation curve we followed the method described in
\citet{2015A&A...573A..59G} 
applied to galaxies of the CALIFA survey. 
First we determined the kinematic center (KC) as the spaxel with zero-rotation velocity. 
We built a distance-velocity diagram for each spaxel with coordinates $x_i$, $y_j$,
where the distance to the kinematic centre ($x_c$, $y_c$) is defined as 
$r_{ij} = \operatorname{sgn}(x_c-x_i)\sqrt{(x_i - x_c)^2 + (y_j -j_c)^2}$, and positive $r_{ij}$ values correspond to increasing values of the 
right ascension. 
Then 
we selected the spaxels with the maximum projected velocity at a given distance $r_{ij}$
to the KC that are expected to trace the kinematic line of nodes   
\citep{1992ApJ...387..503N}.
The kinematic axis position angle (PA$_{\rm kin}$) and the corresponding uncertainty are calculated as the mean and standard deviation of the 
PAs of the radial distance between the selected spaxels and the KC \citep[see][for details]{2015A&A...573A..59G}. 

In VCC~135 the kinematic centre agrees with the position of the optical centre.
We obtained  
PA$_{\rm kin} = 211^{\circ} \pm 11^{\circ}$, and the PAs of the receding and approaching sides 
are comparable within the errors (black squares in the bottom panel of Fig. \ref{fig:vel135}). 
The kinematic and photometric axis show a similar alignment with an angular 
difference of $\Delta\alpha \sim$ 8$^{\circ}$. The orientation of both axis is displayed 
in the top-right corner of the VCC~135 velocity map.
In the top panel of the Fig. \ref{fig:vel135}  we show
the position-velocity diagram of the spaxels included in  
an artificial slit of width 2$^{\prime\prime}$ (2 spaxels) centred on the KC, and orientation defined by the 
measured PA$_{\rm kin}$
\citep[][]{2017MNRAS.470.1991B}. 
The gas velocity of VCC~135 rises linearly  
up to $\sim \pm$ 5\arc\ where it reaches the maximum.  
The velocity profile is fitted with a function $V(r) = V_c (r/r_t)$ 
(i.e. solid-body rotation) for $r < r_t$ and $V(r) = V_c$ for $r > r_t$, where $r_t$ indicates the transition radius 
between the two regimes. We obtain a best-fit radius
 $r_t \sim 3\farcs5$ on both sides. 
The maximum velocity of the receding side  
$V_c^{\rm rec} = 65 \pm 3$ \kms\ (76 \kms, inclination corrected) is higher compared to that of the approaching one, 
$V_c^{\rm app} = -39 \pm 4$ \kms.
Comparison with the stellar velocity map obtained from our data set shows a similar trend with a more symmetric 
receding and approaching maximum velocities 
(filled curve).

VCC~324 displays a much smoother velocity gradient with a more complex kinematics 
(Fig. \ref{fig:vel324}, bottom panel).
The range in velocities across the system is relatively small, 
varying between  $\sim -30$ and $+$20 \kms, with an almost uniform speed in the receding side hovering around $+$10
\kms. At these values possible signatures of rotation become indistinguishable from the
random motion of the gaseous component \citep{2012AJ....143...40M,2016ApJ...832...89M}.
The southern clump, corresponding to the region with an additional metallicity dip in Fig. \ref{fig:met_img}, appears 
to be kinematically disjoint from the surrounding gas as it is moving at $\sim -$20 \kms. 
Given the lack of a clear rotation
pattern it is difficult to identify an obvious kinematic centre from the map. Nonetheless
we apply the method of \citet{2015A&A...573A..59G} to attempt to constrain the gas kinematics.
If we use as KC the zero-velocity spaxel associated to the centre of symmetry of the ionised gas (KC1),
we obtain an average PA$_{\rm kin} = 78^{\circ} \pm 15^{\circ}$ as it is highlighted by the black squares in the bottom panel 
of Fig. \ref{fig:vel324}. This would imply a misalignment of $\Delta\alpha \sim$ 27$^{^\circ}$ between the kinematic and photometric 
major axis (see bars at the top-right corner of the velocity map). However, if we select the zero-velocity spaxel 
in correspondence with the peak of the 
\Ha\ emission (KC2), we obtain a very different result, with a large misalignment between the receding and approaching
sides ($\Delta\alpha \sim 39^{\circ}$; dashed lines in the same panel). Both alternatives though would imply a kinematically disturbed gas disc.
The position-velocity diagram derived in an artificial slit of 2$^{\prime\prime}$ width, centred on KC1, with PA$_{\rm kin} = 78^{\circ}$  
is shown in Fig. \ref{fig:vel324}.
The arctan function $(2V_c/\pi) \arctan(r/r_t)$, where $V_c$ is the 
maximum velocity rotation and $r_t$ the radius where 50\% of $V_c$ is attained, provides a better fit
to the receding-side curve compared to the linear model applied to VCC~135. This function has been successfully applied to 
model rotation curves of local disc galaxies \citep{1997AJ....114.2402C}.
The best-fit asymptotic velocity is $V_c^{\rm rec} = -29 \pm 3$ \kms\ ($\sim -47$ \kms after correcting for an 
inclination $i = $ 38$^{\circ}$). 
On the other hand, the receding-side velocity 
remains almost constant around $11$ \kms\ 
to a radial distance of $\sim 10^{\prime\prime}$ and then increases 
in the outermost spaxels. 

The stellar velocity map of VCC~324 does not present a clear velocity gradient thus it is not displayed in Fig. \ref{fig:vel324}.
As mentioned in Sect. \ref{sec:gals}, the \hi\ component also shows weak evidence of rotation
\citep{1987ApJ...314...57L}, and 
single-dish observations of the 
molecular gas provide a velocity range comparable to the \Ha\ measurements \citep{2016A&A...590A..27G}.

The lack of a regularly rotating disc 
may be due 
to an interaction or merger that is causing an
asymmetry in the gas kinematics as it is observed in other BCDs 
\citep{2001A&A...374..800O,2017AJ....153..132A,2018MNRAS.481..122C}, thus the disc would still be in the process of settling 
into a rotation pattern.
Moreover, 
 misalignments between the kinematic and photometric major axis of the order of what we find in VCC~324 
are usually observed in interacting galaxies 
\citep{2015A&A...582A..21B,2017AJ....153..132A}.  
Simulations predict angles of $\lesssim$ 20$^{\circ}$ in remnants of 3:1 mergers, while 
misalignments larger than $> 30^{\circ}$ are expected in 1:1 mergers
\citep{2007MNRAS.376..997J}. 

\subsection{Analysis of optical images}
\label{subsec:masks}

\begin{figure*}
\begin{center}
\includegraphics[bb=103 253 489 539,width=7.7cm, clip]{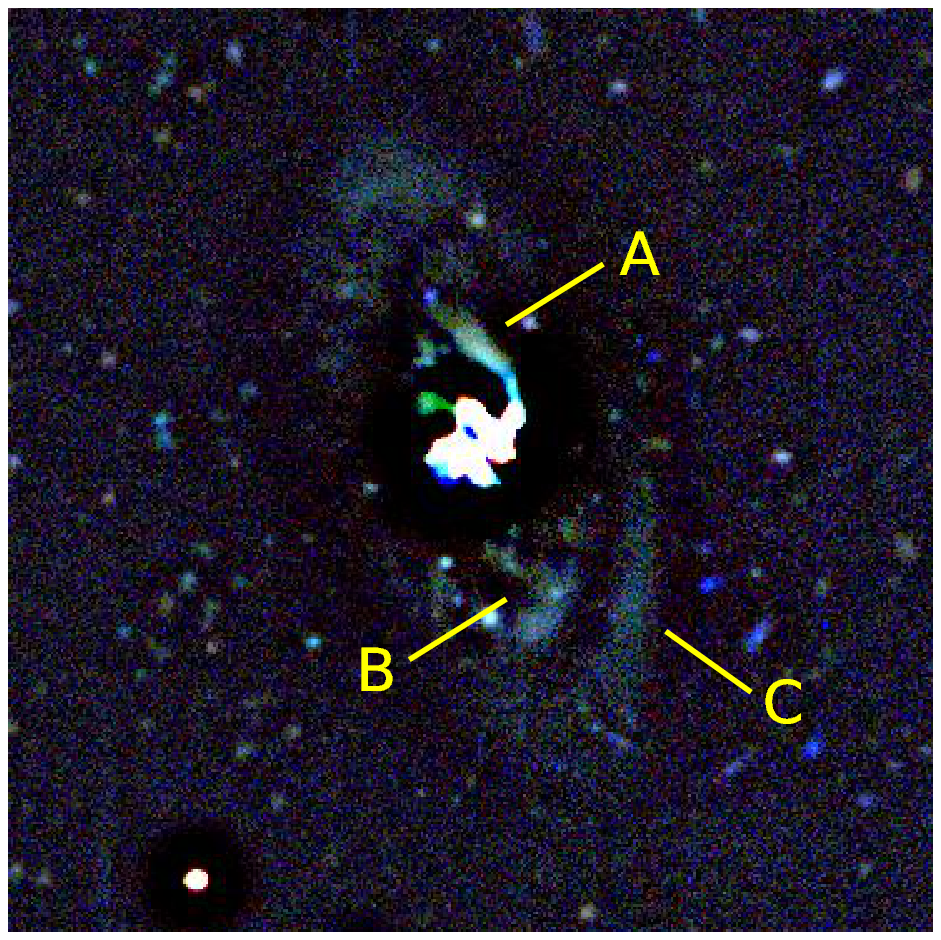}
\advance\rightskip1cm
\includegraphics[bb=113 253 499 539,width=7.7cm, clip]{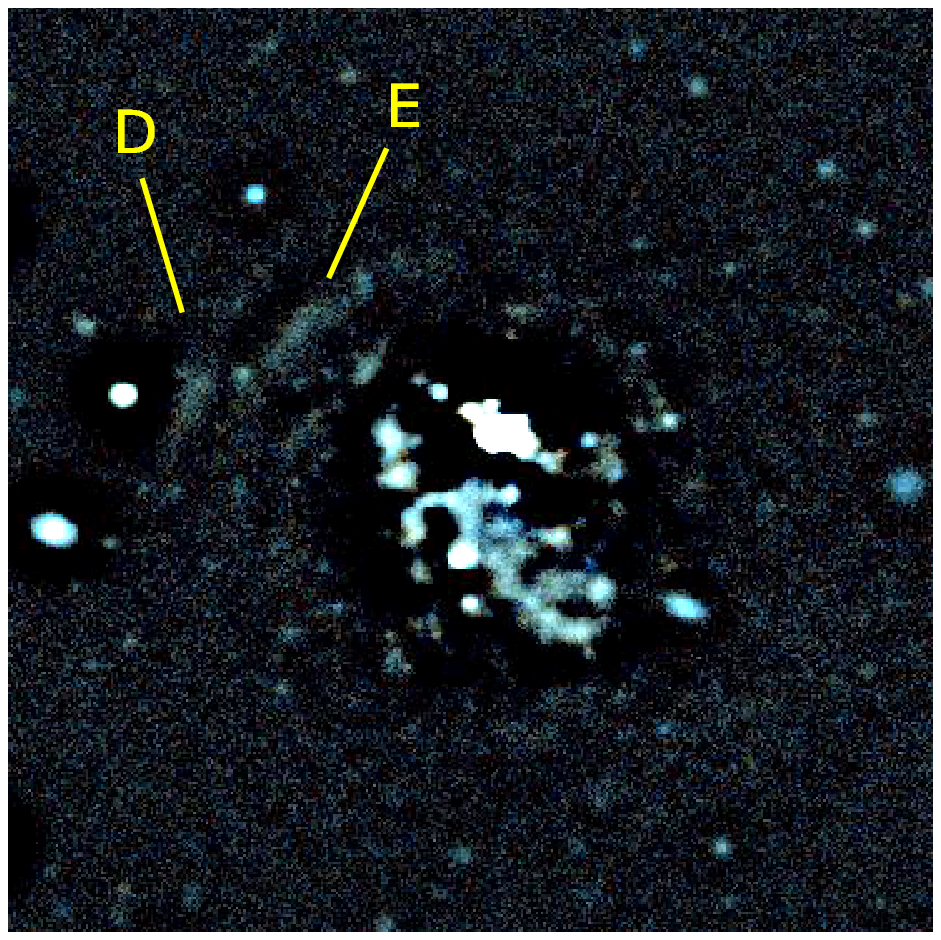}
\includegraphics[bb=-40 -10 655 530,width=7.5cm, clip]{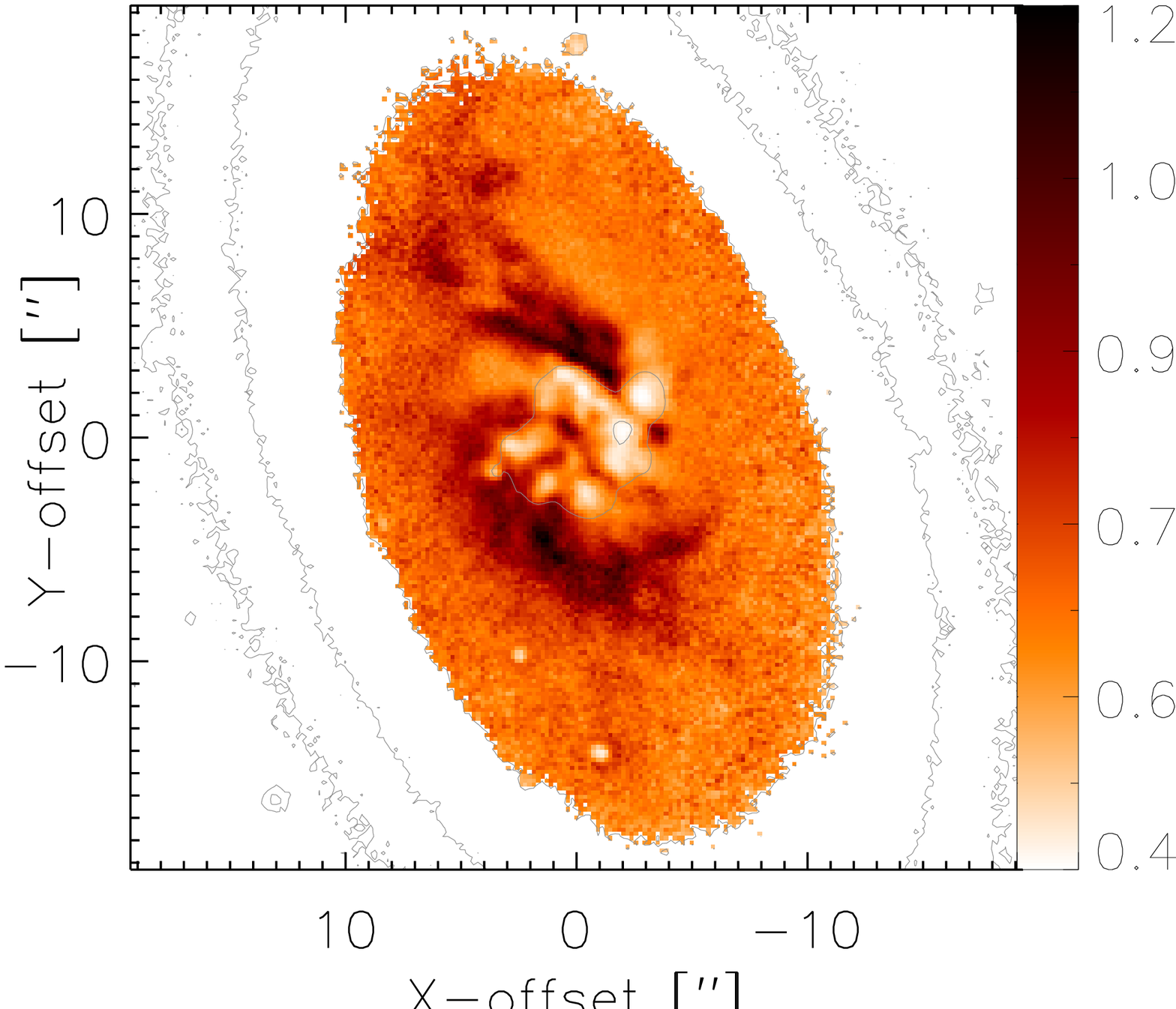}
\includegraphics[bb=-40 -10 655 530,width=7.5cm, clip]{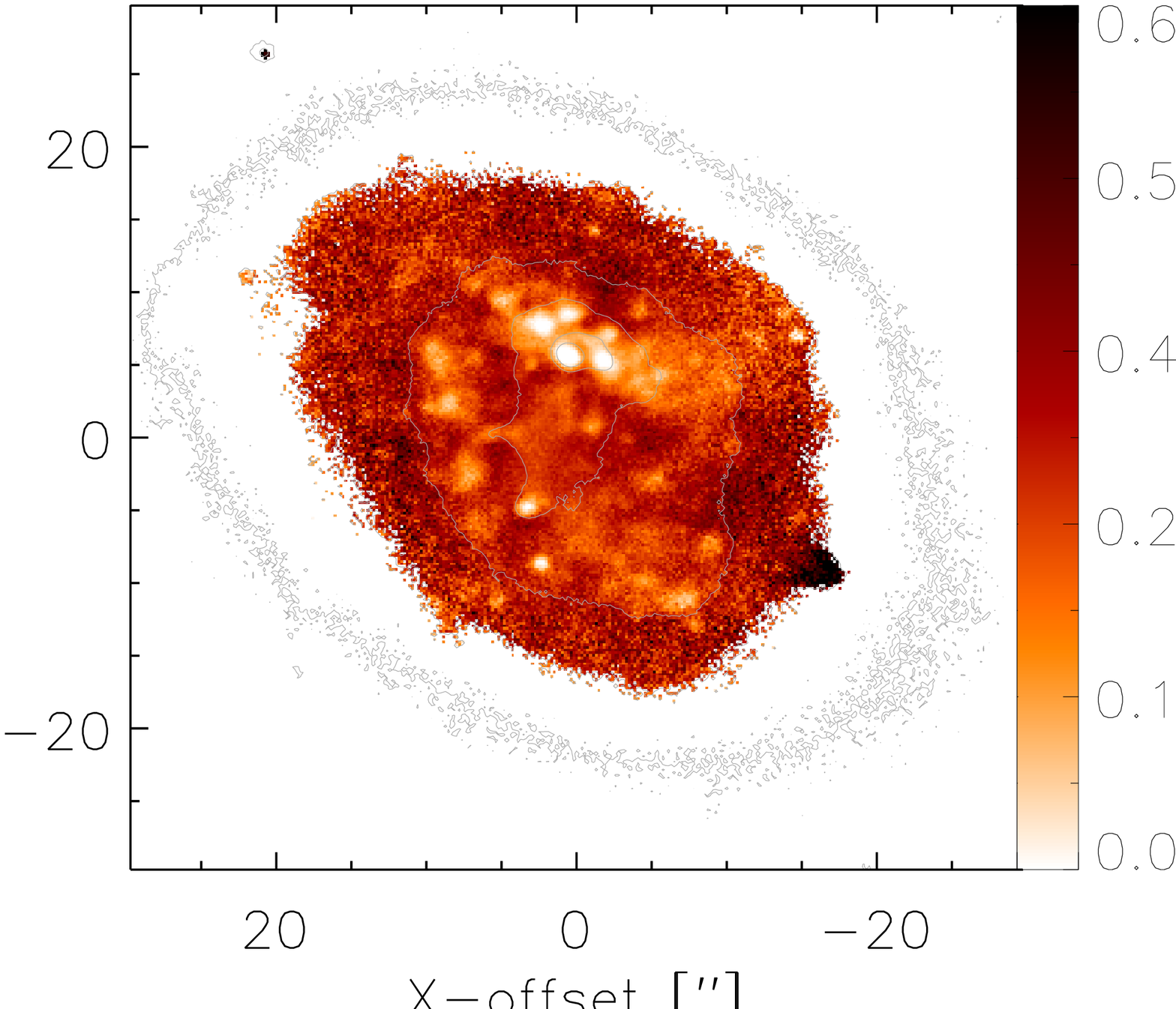}
\caption{Top: Unsharp masks of VCC~135 (left) and VCC~324 (right) with kernel size $\sigma =$ 8 and 10 pixels, respectively.  
Substructures that might indicate a recent merging event are seen in both galaxies and are labeled with letters. 
The cutout size is 75\arc\ (left) and 80\arc\ (right).
Bottom: Colour maps of the inner regions of VCC~135 ($g - i$, left) and VCC~324 ($g - r$, right), respectively.
Red (blue) regions appear as dark (light) areas. North is up and East is left in all panels.}
\label{fig:un_masks}
\end{center}
\end{figure*}

\subsubsection{Unsharp masks and colour index maps}

Unsharp masks are a common technique adopted to
enhance weak substructures like spiral arms, bars, or stellar streams
hidden below the dominant
light distribution \citep{2004A&A...415..941E,2006AJ....132..497L}. 
We produced a set of unsharp masks for each object by
smoothing images obtained from the archive of the NGVCS.  
Images were convolved with a
circular gaussian of various kernel sizes, $\sigma$. Small $\sigma$ values enhance
small structures while large kernel sizes allow the detection of weak large-scale features. 
For each set of unsharp masks
we chose values of $\sigma$ = 2, 3, 4, 5, 6, 8, 10, 12 pixels.
Figure \ref{fig:un_masks} shows the unsharp masks of VCC~135 (left) and VCC~324 (right) with kernel size $\sigma =$ 8, and 10, respectively.
VCC~135 presents a central blue core where most of the star formation is ongoing, with one plume to the north 
(feature A) 
at the edge of the nuclear star-forming region. 
To the south, an S-shaped structure extends from the blue central region 
for about 12\arc (feature B in Fig. \ref{fig:un_masks}). The residual image also shows a ring feature in the outer 
part of the galaxy (C). 
We can rule out that these structures are model-dependent artifacts 
as they also appear in residual images obtained by fitting two S\'ersic components to the $ugiz$ data with \texttt{GALFITM} \citep{2011ASPC..442..479B,2013MNRAS.430..330H}.
The unsharp mask of VCC~324 reveals an elongated jellyfish-shaped structure with several star-forming knots, where most of the ongoing star formation is occurring, as also
shown by the \Ha\ emission. A shell structure (labeled with D and E in Fig. \ref{fig:un_masks}) is visible to the northwest 
side of the galaxy. The shells are more clearly detected in 
the $g$ unsharp mask, suggestive of a relatively young stellar population. 
In both galaxies unsharp masks reveal underlying structure in the stellar discs that appear to trace 
signature of recent interaction/merging events.

Moreover, we built optical colour-index maps to investigate 
the spatial distribution of extinction patches associated to dusty regions\footnote{The color-index 
maps were generated dividing the images
taken in a blue and a red filter ($g - i$, for VCC~135, and $g - r$ for VCC~324) after sky subtraction. 
The seeing of the images in the blue and red filters are very similar, however we convolved
the image with the better seeing with a two-dimensional
Gaussian function with width equal to the quadrature difference
of the two seeing values.}. The bottom-left panel of Fig. \ref{fig:un_masks} shows two red plumes 
($g - i > 0.9$ mag) in VCC~ 135 extending to the north and to the south of the central star-forming region.
The southern plume twist to the west while the norther one extends out to 10\arc\ (approximately 1 kpc at $d = $ 17 Mpc) where two further patchy red features 
are visible.
The plumes may hint at a relic structure originated from the tidal disruption of an accreted satellite galaxy.
For example, the presence of spiral-shaped or irregular dust morphologies in early-type 
galaxies has been interpreted as evidence of 
gas-rich minor mergers \citep{1995AJ....110.2027V,2006ApJS..164..334F,2020arXiv200108087Y}.
The $g - r$ map of VCC~ 324 shows the inner jellyfish-shaped bluer region associated to ongoing star formation. 
There is not a clear dusty structure as in VCC~135
although darker patchy regions are visible in between the trail of bright clusters and blue associations.
In both systems the blue star-forming regions show a different spatial distribution than the outer structure of 
the galaxy associated to the older stellar populations.

\subsubsection{CAS and Gini-$M_{20}$ diagnostics}

As a further test we analysed the $g$-band images of both galaxies with \texttt{statmorph} \citep{2019MNRAS.483.4140R},
a Python package for calculating non-parametric morphological diagnostics such as the Gini coefficient ($G$),
the second moment of the brightest pixels of a galaxy containing 20\% of the total flux  
\citep[$M_{20}$;][]{2004AJ....128..163L}, and the concentration-asymmetry-smoothness system \citep[CAS; see][for details]{2003ApJS..147....1C}. 
These diagnostics have been extensively applied to quantify galaxy morphologies, also allowing to identify systems 
with signatures of recent or ongoing mergers.
The asymmetry index ($A$) is obtained by subtracting the galaxy image rotated by 180$^{\circ}$ from the original one,
and dividing the sum of the absolute value of the residuals to the original galaxy's flux \citep{1996ApJS..107....1A}.
Galaxies with $A \gtrsim 0.2$ are considered morphologically disturbed and those with
$A \geq 0.35$ are likely to be major-merger remnants \citep{2000ApJ...529..886C,2003ApJS..147....1C,2007ApJ...666..212D}.
Originally used in economics to quantify the unequal distribution
of wealth in a population, the Gini coefficient is adopted in astronomy to provide a measure of the
relative distribution of light within the galaxy image \citep{2003ApJ...588..218A}. $G$ $=$ 0 means an equal distribution
across all galaxy pixels, while $G =$ 1 corresponds to extreme inequality where all the light is concentrated within a few
pixels. The position of a galaxy in the $G - M_{20}$ plane allows us to separate major mergers from non-interacting
galaxies \citep{2004AJ....128..163L}. Mergers occupy the region of the parametric space defined by the following relation:
$S(G, M_{20}) = 0.14 M_{20} + G - 0.33 > 0$ \citep{2008MNRAS.391.1137L}.
We ran \texttt{statmorph} on the $g$ images and derived the diagnostics  discussed above for the two galaxies that we 
display in Table \ref{tab:morph}.
In VCC~324  the asymmetry index $A$ is above the threshold value that defines major-merger candidates, while its $S(G, M_{20})$ 
places it at the very edge of the region that separates mergers from ''normal`` irregular galaxies in the $G - M_{20}$ plane. 
On the other hand, VCC~135 with $S(G, M_{20}) \sim -0.08$ and $A = 0.14$ 
is below the canonical major-merger separation thresholds.
However this does not rule out that VCC~135 might have experienced a minor merger event. 
\citet{2012MNRAS.419.2703H} report that in late minor mergers in which the
less massive galaxies have been almost entirely dissolved -- as the analysis of VCC~135 images may suggest -- the application
of these diagnostics is less effective. 

\begin{table}
\begin{center}
\begin{tabular}{lcccccc}
\hline \hline
ID      & $C$  & $A$  & $S$  & $G$ &$M_{20}$& $S(G, M_{20}$) \\ 
\hline \hline
VCC~135 & 3.62 & 0.14 & 0.11 & 0.54 & -2.09  &  -0.08 \\
VCC~324 & 2.69 & 0.36 & 0.04 & 0.54 & -1.43  &   0.01\\
\hline \hline
\end{tabular}
\caption{CAS coefficients and Gini $-$ $M_{20}$ statistics from the \texttt{statmorph} package.}
\label{tab:morph}
\end{center}
\end{table}

\section{Discussion}
\label{sec:discuss}

The observed anticorrelation between the metallicity of the
gas and the SFR is expected in a scenario where the star-formation process is triggered by  
metal-poor gas falling onto the disc. 
Two mechanisms can cause the inflow of metal-poor gas: accretion from the IGM
 \citep{2016MNRAS.457.2605C} 
or galaxy-galaxy interactions/mergers \citep{2012ApJ...753....5R}.
Alternatively, intense outflows produced by stellar winds and supernovae (SNe) could 
eject metals from the central star-forming regions, diluting the heavy-element abundance \citep{2004ApJ...613..898T,2014A&A...563A..58T}.

Models of galaxy formation predict that gas accretion
from the cosmic web is a primary driver of star formation at early epochs \citep{2013MNRAS.435..999D}, and that at dark matter halo masses of M$_h = 10^9 - 10^{10}$ \msun\ cold gas accretion still
 occurs in low-density environments and filaments at $z$ = 0 \citep{2008A&ARv..15..189S,2009MNRAS.395..160K}.
Analytical models of \citet{2013MNRAS.435.2918M} are able to reproduce the observed inverted gradients in high
redshift galaxies introducing a high inflow rate of pristine gas towards the inner region of their discs.
In the external gas accretion scenario the metallicity of the infalling gas must be much lower than solar 
 \citep{2016ApJ...819..110S,2018MNRAS.476.4765S}; the accreted gas would
mix in a time-scale of the order of the rotational period, averaging the inverted gradient in a few hundred Myr \citep{2012ApJ...758...48Y,2015MNRAS.449.2588P}.  

Extended \hi\ discs with disturbed morphologies and kinematics are thought to be associated with intergalactic gas flows 
\citep{2008A&ARv..15..189S,2011AJ....141....4K}.
The VLA map of VCC~324 shows a \hi\ extension of about 
1$^{\prime}$ at a column density of 1.5 $\times 10^{20}$ cm$^{-2}$  \citep{1987ApJ...314...57L}, typical of late-type dwarfs. It
is difficult to confirm or reject the external gas accretion scenario from these data,
requiring deeper  aperture synthesis observations.
High-resolution 21-cm data of VCC~135 are not available in the literature.
Accretion of gas from the IGM would be suppressed in a dense environment, however neither galaxy is
located in the very central regions of Virgo. Adopting a distance of 17 Mpc, 
VCC~135 and VCC~324 are at about 1.1 and 1.4 Mpc from M87 (cluster A) and M49
(cluster B), respectively. Both observations and simulation suggest 
that galaxies in the periphery of clusters ($r \gtrsim 2.4$ Mpc) can accrete cool gas from their surroundings. 
However, at smaller distances to the cluster centre, starvation (1 Mpc $< r <$ 2.4 Mpc) and ram pressure 
stripping ($r < 1$ Mpc) can inhibit this process \citep{2007ApJ...671.1434T,2015PKAS...30..495Y}.

Mergers and interactions between dwarf galaxies have been proposed to explain the properties of starbursting BCDs
\citep{2001A&A...373...24P,2008MNRAS.388L..10B}.
In this scenario, metal-poor gas from the outskirts of the progenitor discs is brought towards the centre by 
the interaction with another gas-rich dwarf, 
similarly to what is found in more massive galaxy mergers \citep{2012ApJ...753....5R}. 
\citet{2018MNRAS.474.2039E,2020MNRAS.492.6027E}
found that interacting 
starburst galaxies (M$_* > 10^{10}$ \msun) present lower central metallicities
suggesting that metal-poor gas inflows from mergers can
dilute the oxygen abundance. 
On the other hand, \citet{2015A&A...579A..45B} determined that the central metallicities of interacting systems 
extracted from the CALIFA survey are comparable to those of a non-interacting control sample, suggesting that stellar feedback 
could be responsible for enriching  the ISM in the center\footnote{The apparently contradicting results of the two studies may be due to the different criteria adopted to select 
the interacting/merging galaxy samples: \citet{2015A&A...579A..45B} included systems at 
varied stages of interaction (from pre-merger pairs with separations as large as 150 kpc to post-mergers) while 
\citet{2018MNRAS.474.2039E} considered only galaxies above the main sequence with enhanced star formation activity.}. 

Simulations of massive galaxies give evidence that galaxy-galaxy interactions can flatten the metallicity gradient, 
but it is not clear whether they are capable of inverting it 
\citep{2017MNRAS.472.4404S,2018MNRAS.478.4293C}. 
Central metallicity depressions of the order of $\lesssim$ 0.2
dex are predicted \citep{2012ApJ...746..108T}, and such values are in agreement with observations 
\citep{2008ApJ...674..172R,2013MNRAS.435.3627E}.
Models of low-mass discs interactions show for example that 
galaxies with stellar masses of M$_* \sim 10^9$ M$_{\odot}$ and
initial nuclear metallicities of 12 + log(O/H) $\sim$ 8.6 could attain 
12 + log(O/H) $\sim$ 8.3 or 8.4 in the centre during
a merger \citep{2012ApJ...746..108T}. If low-mass galaxies are on average characterised by flat metallicity 
gradients \citep{2012MNRAS.427..740L,2017MNRAS.464..739M,2017MNRAS.469..151B,2019IAUS..344..161G} the central depression combined with  a flat distribution at larger 
radii could produce an overall positive trend 
similarly to what we observe in our targets.

The results that we presented in Sect. \ref{sec:res} 
seem to support the merger scenario. The lack of a clear rotation gradient and the kinematic axis misalignment in VCC~324 may imply that the galaxy went through a major
merger (mass ratio $\gtrsim$ 3:1) as it is also may be suggested by the Asymmetry parameter. 
Shell-like stellar structures similar to what are found in VCC~324, are associated with
intermediate-mass/major mergers \citep{2018MNRAS.480.1715P} and they have been observed in both massive 
\citep{2015MNRAS.446..120D,2018ApJ...866..103K} and dwarf galaxies \citep{2017ApJ...834...66P,2020ApJ...891L..23Z}.
For what concerns VCC~135, analysis of the gas kinematics and of the morphology indicators would exclude a major merger.
However, the structures that we find in the unsharp masks and colour index maps give hints of an interaction event
possibly related to the accretion of a lower-mass satellite. Given the high \hi\ deficiency of this galaxy it is plausible
that most of the detected gas is associated to the accreted object and under this assumption we can estimate an upper limit
to the mass of the disrupted dwarf. 
For a total gas mass  M$_{gas} = 4 \times 10^7$ \msun (including H$_2$ and \hi, Table \ref{tab:prop}),  
assuming that the accreted object is gas-rich \citep[M$_{gas}$/M$_* \sim 1$;][]{2012ApJ...756..113H},
comparison with VCC~135 stellar mass  
would imply that the galaxy may have experienced 
a minor merger event with ratio $<$ 1:10.

\begin{figure}
\begin{center}
\includegraphics[bb=50 20 540 450,width=8.5cm, clip]{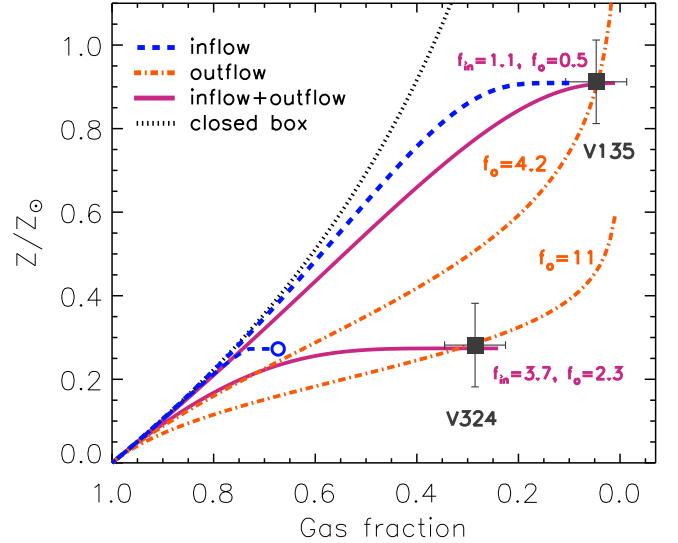}
\caption{Evolution of metallicity with gas fraction for simple chemical evolution models with pure outflows 
(dash-dotted orange lines), pure inflows (dashed blue lines), and a combination of inflows and outflows 
(solid purple lines), compared to the observed values of  VCC~135 and VCC~324 (filled squares). 
$f_{in}$ and $f_o$ show the assumed inflow and outflow rates in units of the galaxy SFR. Pure accretion models have the 
same $f_{in}$ as the mixed ones. The black dotted line displays
the expected trend from a closed-box model.}
\label{fig:met_model}
\end{center}
\end{figure}

Lastly, outflows due to SN feedback ejecting enriched gas from the inner regions 
could also produce the observed metallicity trend. 
This process is found to be effective in galaxies at high redshift \citep{2019ApJ...882...94W} 
with log(sSFR/yr$^{-1})$ = -7.7, well above
those observed in our targets (by a factor of 1.8 dex in VCC~324 and 2.8 dex in VCC~135, respectively; see Table \ref{tab:prop}).
To test this possibility, in Fig. \ref{fig:met_model} we compare the properties of our galaxies with 
the simple chemical evolution model of \citet{2008ApJ...674..151E}.
The model has two free
parameters: the mass loading factor, $f_o = \dot{M}_{out}/$SFR, and the mass accretion factor,
$f_{in} = \dot{M}_{acc}/$SFR, where $\dot{M}_{out}$ and $\dot{M}_{acc}$  are the mass-loss and mass-accretion rates, 
respectively\footnote{The model adopts the
instant recycling and mixing approximations. The metallicity of the infalling gas is assumed
to be zero, while the outflows have the same abundance as the galaxy ISM, and  
the fraction of gas returned to the ISM by star formation is neglected.}.
In a pure-outflow scenario 
high mass loading factors ($f_o  =$ 4.2 and 11) would be needed to reproduce the observed gas fraction and metallicities of
the two dwarfs (Fig. \ref{fig:met_model}). 
Observations of nearby starburst dwarfs 
reveal that there are relatively
few strong outflows in low-mass galaxies in the local Universe \citep{1999ApJ...513..156M,2019ApJ...886...74M}. 
The measured mass loading factors range between 0.2 and 7 with a weak dependence
on stellar mass ($\propto$ M$_*^{0.04}$), and only few objects have $f_o$ above 2. 
An average $f_o \sim$ 1.2 is expected for a galaxy with stellar mass M$_* \sim 10^9$ \msun \citep{2019ApJ...886...74M},
in agreement with results 
from hydrodynamical simulations at this mass range \citep{2016ApJ...824...57C}. 
Given the relatively low sSFR of our targets and the high mass loading factors required by pure outflow scenarios, 
it appears unlikely that stellar feedback only is able to cause the observed inverted 
metallicity gradients. 

A pure inflow model with mass-accretion factor $f_{in}$ = 
1.1 (dashed blue line, Fig. \ref{fig:met_model}) or a combination of 
$f_{in} = 1.1$ and a low outflow rate ($f_{o} = 0.5$, solid blue line) appear more favoured in VCC~135. 
For VCC~324 we find a best-fitting model with $f_{in} = 3.7$ and $f_{o} = 2.3$,
while a pure inflow scenario with the same $f_{in}$ (dashed line) is not able to reproduce the observed galaxy properties. 
This is because the model predicts that in the absence of outflows 
the gas reservoir cannot be completely exhausted, thus 
$f_{gas}$ cannot decrease beyond a 
minimum threshold set by the inflow rate, $f^{min}_{gas} = 1 - 1/f_{in}$ 
(empty dot in Fig. \ref{fig:met_model}).
Even though outflows may be present in these systems this analysis suggests that they do not 
dominate the chemical evolution of our targets, that is more likely driven by the infall of low-metallicity gas possibly related to a recent
interaction. The lower mass accretion factor obtained for VCC~135 compared to VCC~324 would support our interpretation of a minor- versus 
major-merger event.

\section{Summary and Conclusions}
\label{sec:concl}

We presented integral field spectroscopy observations 
of two star-forming dwarf galaxies in the Virgo cluster obtained with PMAS/PPak at the Calar Alto 3.5 meter telescope. 
We derived metallicity maps using the N2 indicator. 
The galaxies show inverted metallicity gradients, contrary to what is usually observed in dwarfs or in spiral galaxies.
We find gradient slopes of 0.20 $\pm$ 0.06 and  
0.15 $\pm$ 0.03 dex/$R_e$ for VCC~135 and VCC~324, respectively. The slopes are steeper than what is found in previous
studies of local spirals or high-$z$ dwarfs with similar positive gradients.
We discussed whether such a trend could be caused by inflow of metal-poor gas -- accreted from 
the IGM or in a recent merging event -- or to enriched-gas outflows triggered
by the star formation activity. Comparison with simple chemical evolution models seem to favour the gas-inflow scenarios.
However their location in the outskirts of the Virgo cluster would hinder external gas accretion from the IGM
due to the effects of ram-pressure stripping and/or starvation.
Analysis of deep optical images and of the ionised gas kinematics suggest that both 
galaxies may have recently accreted a gas-rich companion. 
We argue that VCC~324 is the remnant of a major merger, while a minor-merger
scenario seems more favoured for VCC~135. 
The accretion event drove metal-poor gas from the
galaxy outskirts to the central regions, causing the inverted metallicity trend across the galaxy discs.
Atomic hydrogen maps are needed to investigate the
\hi\ distribution in the two systems in order to further constrain our interpretation of a 
dwarf-galaxy merging scenario.

\section*{Acknowledgements}

The authors wish to thank the Calar Alto telescope staff for help and support during the observing runs.
We thank the anonymous referee for his/her suggestions and comments that contributed to improve the manuscript.
RGB acknowledges support from the Spanish Ministerio de Econom\'ia y
Competitividad, through project AYA2016-77846-P and by the
Spanish Science Ministry "Centro de Excelencia Severo Ochoa Program
under grant SEV-2017-0709. PAAL thanks the support of CNPq, grant 309398/2018-5.
We acknowledge the usage of 
\texttt{MPFIT} \citep{2009ASPC..411..251M} routines.
This work is based on observations obtained with
MegaPrime/MegaCam, a joint project of CFHT and CEA/
DAPNIA, at the Canada--France--Hawaii Telescope (CFHT),
which is operated by the National Research Council (NRC)
of Canada, the Institut National des Sciences de l'Univers of
the Centre National de la Recherche Scientifique (CNRS) of
France, and the University of Hawaii.

\section*{Data availability statement}

The data underlying this article will be shared on reasonable request to the corresponding author.




\bibliographystyle{mnras}
\bibliography{IFUbib} 




\appendix

\section{Alternative estimates of the oxygen abundance of VCC~324}
\label{sec:app}

In this section we 
discuss alternative methods to determine the gas-phase metallicity. 
As mentioned in Sect. \ref{subsec:metal} the N2 ratio depends on
physical parameters of a nebula other than the metallicity.  
Therefore 
variations of the N2 index could be correlated, for example, to changes in the ionisation parameter\footnote{The ionisation parameter, defined as the ratio of ionising photon density to hydrogen density, 
is sometimes defined in the literature in the dimensionless form $U = q/c$ where $c$ is the speed of light.}
($q$) instead of metal abundance \citep{2012MNRAS.426.2630L,2014ApJ...797...81M}.
On the other hand, \citet{2018MNRAS.476.4765S} showed that the observed anticorrelation between N2-based metallicities and SFR 
in a sample of star-forming dwarfs still holds when O/H is calculated with photoionisation models that take into account
the variation of the physical properties of a nebula. 
It is thus important to test whether different metallicity estimates can reproduce the results that we derived with the N2 calibrator.

\begin{figure}
\begin{center}
\includegraphics[bb= -10 -10 690 530,width=8.7cm, clip]{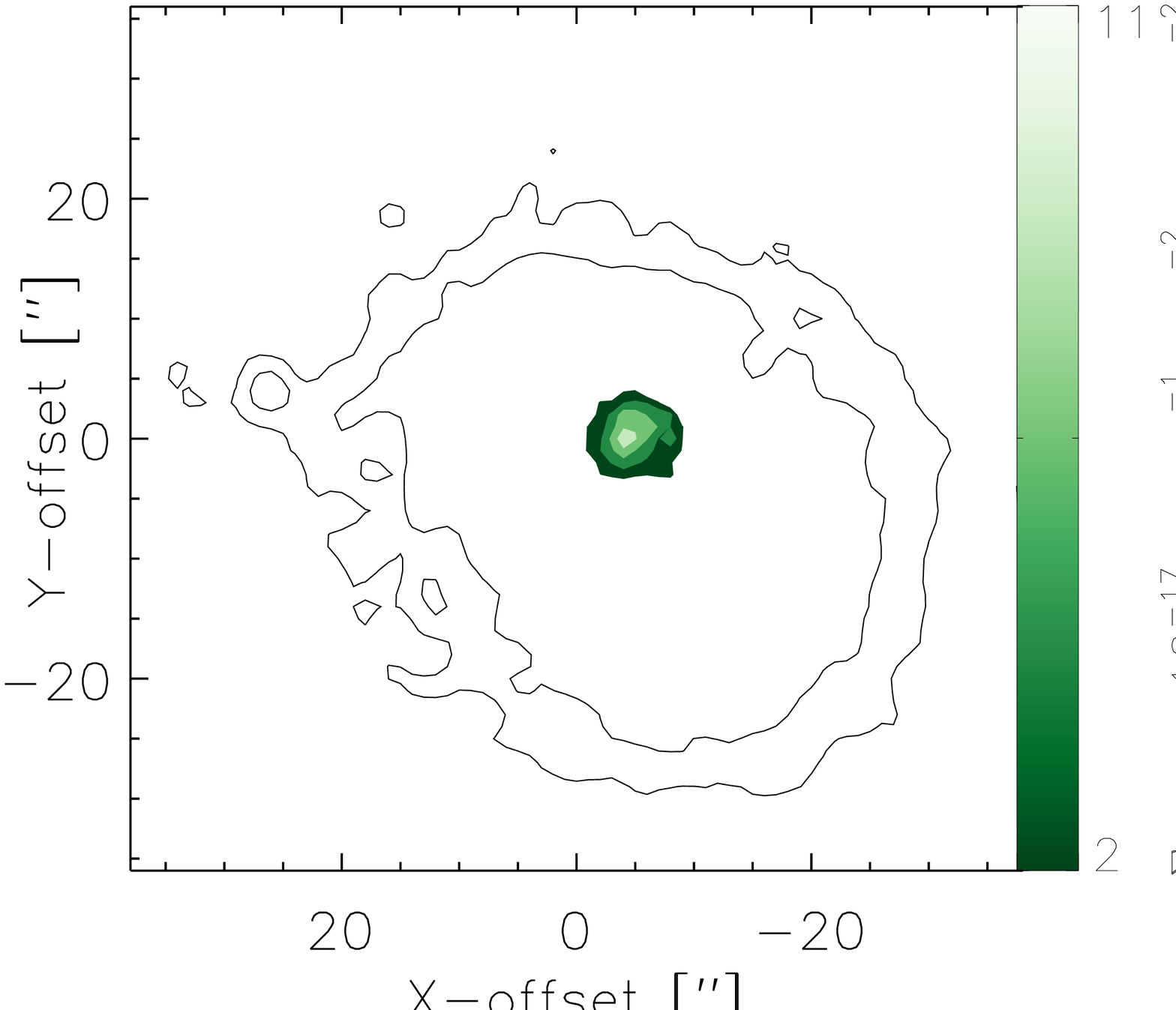}
\caption{ \oiii$\lambda$4363 map of VCC~324. 
    Contours range between 2.2 and 11 $\times$ 10$^{-17}$ \cgs\ arcsec$^{-2}$. The lowest value
    is three times the rms and the contours indicates 3,7,10,15$\sigma$.}
\label{fig:app:oiii}
\end{center}
\end{figure}

An alternative way to determine O/H is provided by the direct method 
\citep{1954ApJ...120..401A,1990ASSL..161..257D,2006agna.book.....O},
based on the detection of the auroral line \oiii$\lambda$4363 that allows to measure the 
electron temperature $T_e$(\oiii).
Assuming a two-zone approximation \citep{1992AJ....103.1330G}, 
the electron temperature $T_e$(\oii) in the O$^+$ zone is measured from 
weak auroral lines such as 
\oii$\lambda\lambda$7320,7330, or 
it is calculated from $T_e$(\oiii) through a relation
derived from photoionisation models \citep{1992AJ....103.1330G,2014MNRAS.441.2663P}.
The metallicity is then obtained from 
the relations linking the oxygen ionic abundances (O$^+$/H, O$^{++}$/H) to the corresponding electron temperatures and observed
emission-line ratios 
\citep{2006agna.book.....O}.
However, auroral lines are intrinsically faint and thus more difficult to observe. 
In VCC~324 \oiii$\lambda$4363 is detected above 3$\sigma$ only in 
a very small region  ($r < 0.6 R_e$; Fig. \ref{fig:app:oiii}), 
and \oii$\lambda$7320,7330 is observed in an even more compact area, smaller by a factor of two.
The reduced extension of \oiii$\lambda$4363 emission compared to the size of the galaxy, its relatively low S/N ratio 
(only a few spaxels have S/N $>$ 10),
the lack of the 
auroral lines 
that allows to more accurately determine
$T_e$(\oii) and the O$^+$ abundance \citep{2020A&A...634A.107Y}, hamper the application of this method.

\begin{figure*}
\noindent
  \subfloat{  
  \includegraphics[bb= 0 -15 560 620,width=0.34\textwidth, clip]{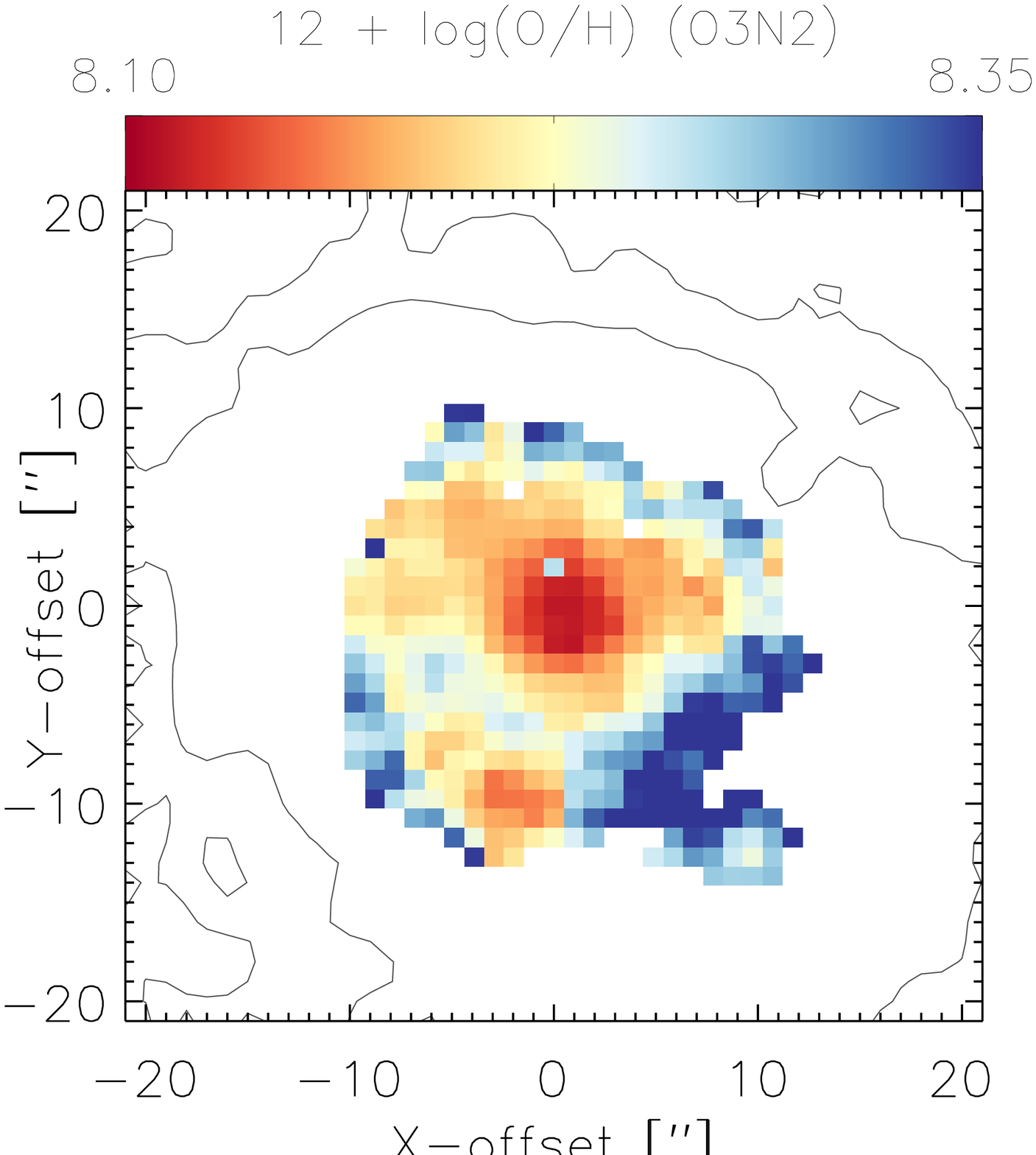}} 
  \subfloat{ 
  \includegraphics[bb= 0 -15 560 620,width=0.34\textwidth, clip]{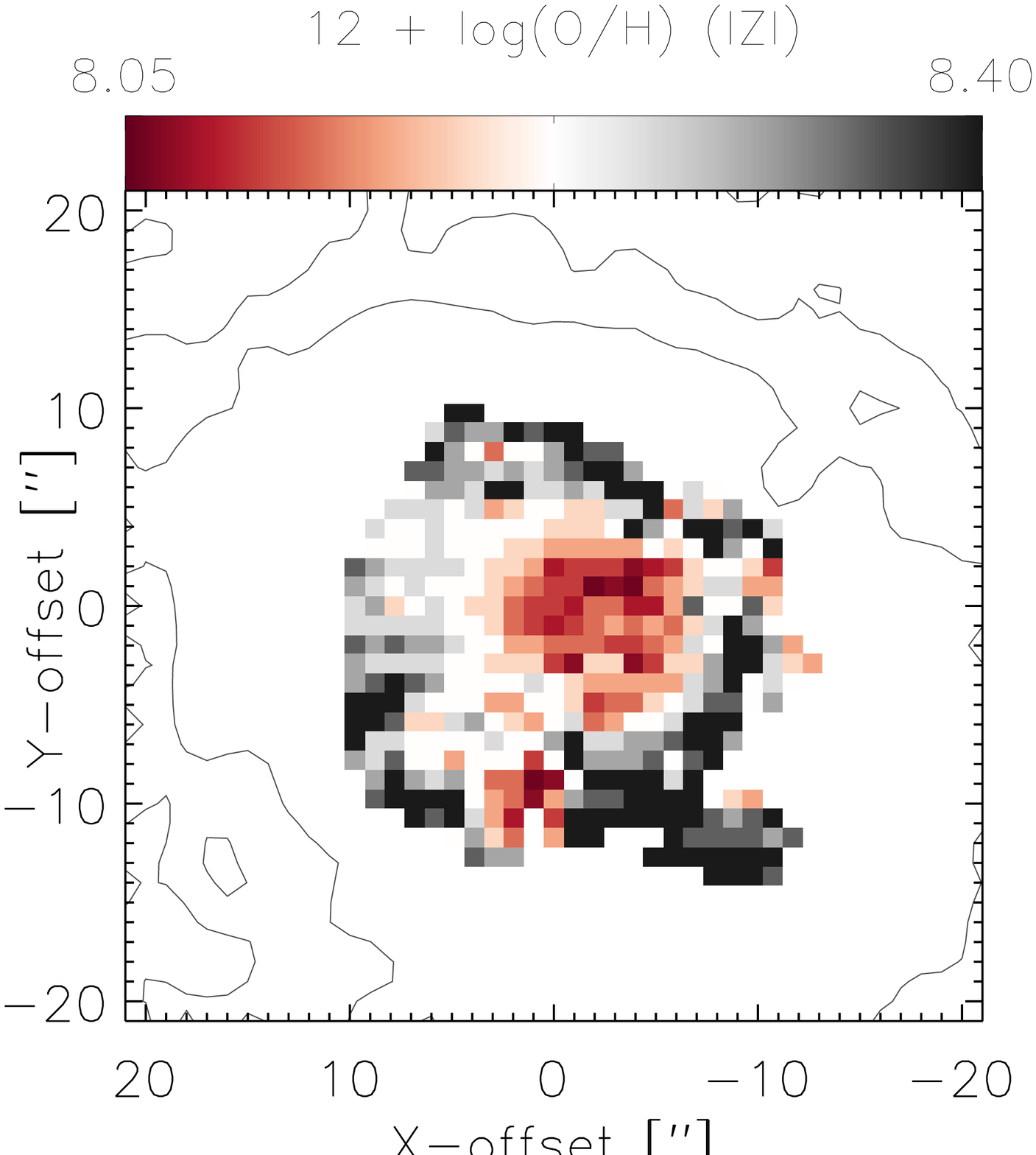}} 
  \subfloat{ 
  \includegraphics[bb= 0 -15 560 620,width=0.34\textwidth, clip]{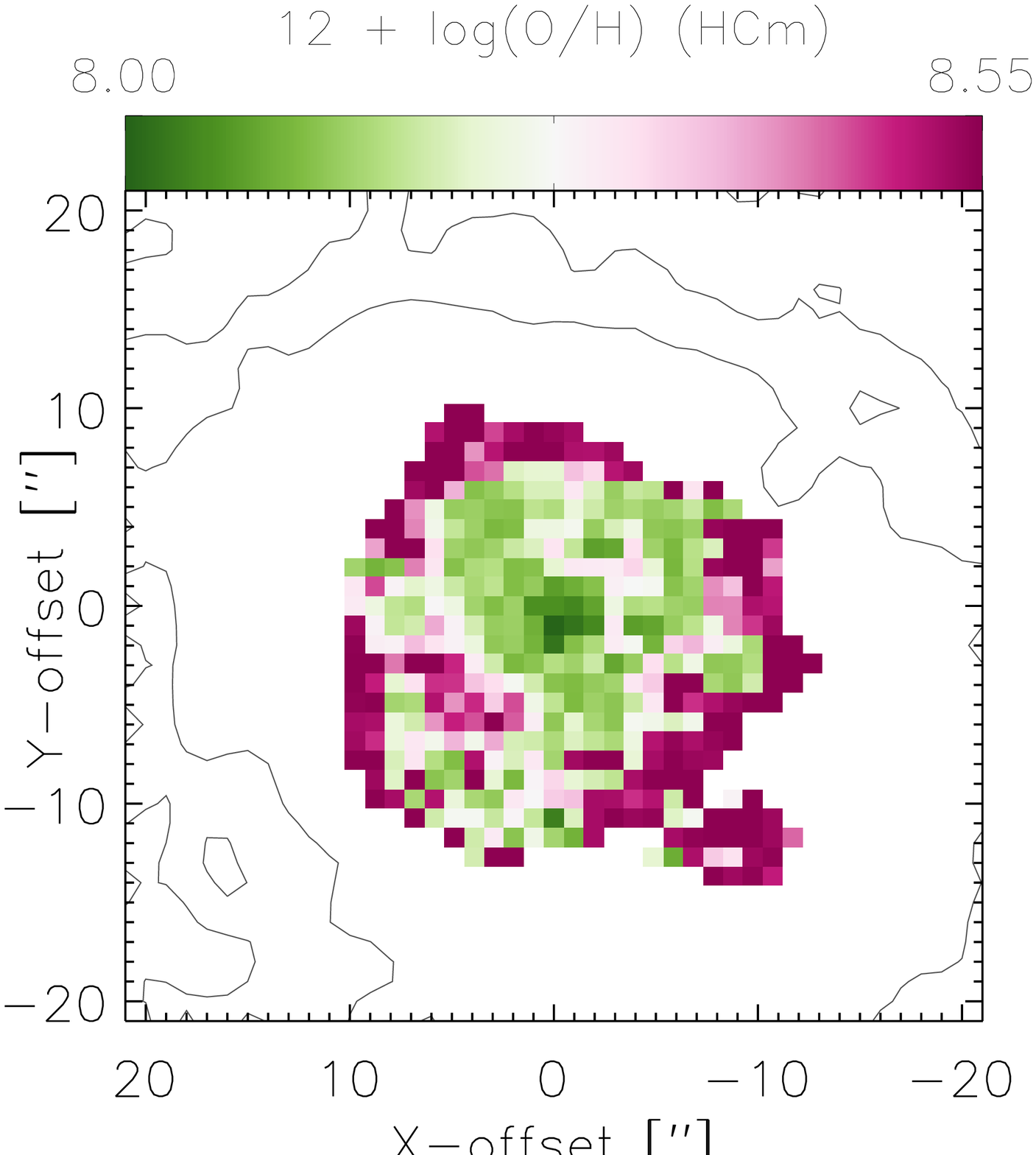}} 
    \caption{Oxygen abundance maps of VCC~324 obtained with the O3N2 calibrator (left), \texttt{IZI} (centre),
    and \texttt{HCm} (right). The metallicity estimate is performed on a smaller region compared to Fig. \ref{fig:met_img}
    to ensure that the line fluxes included in the analysis have S/N $\gtrsim$ 10.}
    \label{fig:app:met_maps}
\end{figure*}


Therefore, here we test another widely-used
empirical calibrators, O3N2 (see Sect. \ref{subsec:metal}), and 
metallicity diagnostics based on photoionisation models: 
i) Inferring metallicities ($Z$) and Ionisation parameters  
\citep[\texttt{IZI}\footnote{
\url{https://users.obs.carnegiescience.edu/gblancm/izi/}};][]{2015ApJ...798...99B}; ii) 
\texttt{\hii-CHI-MISTRY}\footnote{
\url{https://www.iaa.csic.es/~epm/HII-CHI-mistry-opt.html}} 
\citep[\texttt{HCm, v4.1};][]{2014MNRAS.441.2663P}. The comparison is only performed for VCC~324 because it is the target
where line emissions cover a larger fraction of the galaxy disc, allowing to better investigate the spatial variation of the 
metallicity.

The O3N2 map is displayed in the left panel of Fig. \ref{fig:app:met_maps}.
In the figure we are showing a smaller region compared to Fig. \ref{fig:met_img} to ensure that
we are analysing only spaxels where the observed strong-line fluxes have S/N $\gtrsim$ 10. 
The map presents a similar range of O/H and spatial trend to N2. 
However this could be due to the fact that O3N2 is also
sensitive to both metallicity and ionisation parameter \citep{2002ApJS..142...35K,2010IAUS..262...93S}. 

\texttt{IZI} provides a method to determine the physical properties of a nebula from strong emission lines 
without using a specific metallicity indicator.
The tool applies Bayesian inference to calculate the joint and marginalised probability
density function (PDF) of the oxygen abundance (and $q$) for a given combination of observed lines,
comparing simultaneously all the available line ratios to the predictions of three
different sets of photoionisation models 
\citep{2001ApJ...556..121K,2010AJ....139..712L,2013ApJS..208...10D}. 
A uniform maximum ignorance prior in $q$ and $Z$ is assumed.
The procedure allows to include different strong emission lines as input, 
and we used the following set tested in \citet{2015ApJ...798...99B}:
\oii$\lambda\lambda$3727,3729,
H$\beta$, \oiii$\lambda$5007, \Ha, \nii$\lambda\lambda$6548,6584, [S {\sc ii}]$\lambda\lambda$6717,6731.
Line fluxes were corrected for extinction using the Balmer decrement assuming case B recombination and applying a \citet{1989ApJ...345..245C} 
galactic extinction law.
The \texttt{IZI} output, shown in the central panel of Fig. \ref{fig:app:met_maps}, was obtained using
the \citet{2001ApJ...556..121K} grids with 12 + log(O/H)$_{\odot}$ = 8.69 \citep{2009ARA&A..47..481A};
qualitatively similar results are found with the other photoionisation models implemented in the procedure.
In this case,
the input ionising spectrum is computed with \texttt{STARBURST99} \citep{1999ApJS..123....3L} 
assuming a \citet{1955ApJ...121..161S}
initial mass function (IMF) and an age of 8 Myr for the stellar population. 
The \texttt{IZI} output provides a measure of the metallicity which is consistent with O3N2 and N2, 
displaying a central dip and an abundance increase with radius. 

\texttt{HCm} determines chemical abundances (O/H, N/O) and the ionisation parameter 
performing a $\chi^2$ minimization procedure between the observed line fluxes and the predictions of a  grid of models
calculated with \texttt{CLOUDY} \citep[\texttt{v17.00};][]{2013RMxAA..49..137F}.
The ionising radiation field is simulated with the \texttt{POPSTAR} code 
 \citep{2009MNRAS.398..451M} assuming a starburst age of 1 Myr and a \citet{2003PASP..115..763C} IMF. 
The procedure first determines the N/O ratio using
strong lines 
such as \nii$\lambda$6584/\oii$\lambda$3727 and 
\nii$\lambda$6584/[S {\sc ii}]($\lambda$6717 + $\lambda$6731).
Once N/O is constrained, the oxygen abundance and the ionisation parameters are determined in a new iteration 
using other combinations of the available line ratios \citep[see][for details]{2014MNRAS.441.2663P}.
When the \oiii$\lambda$4363 line is lacking an empirical relation between O/H and $q$ is assumed.
\texttt{HCm} allows to use the same set of line fluxes as \texttt{IZI}
with the inclusion of [Ne {\sc iii}]$\lambda$3869 and \oiii$\lambda$4363, if detected.
We ran \texttt{HCm} with [Ne {\sc iii}]$\lambda$3869
among the list of observables and its output is displayed in the
right panel of Fig. \ref{fig:app:met_maps}. 
Including [Ne {\sc iii}] results in a smoother increase of the metallicity with radius 
from 12 + log(O/H) $\sim$ 8.1 to 8.5 dex. Compared to the other methods \texttt{HCm} measures a higher abundance
at the edge of the displayed region. 
The advantage of this procedure is that it allows to derive
chemical abundances that are consistent with the direct method
\citep{2014MNRAS.441.2663P}.

\begin{figure}
\begin{center}
 \includegraphics[bb= -20 -3 580 585,width=8.7cm, clip]{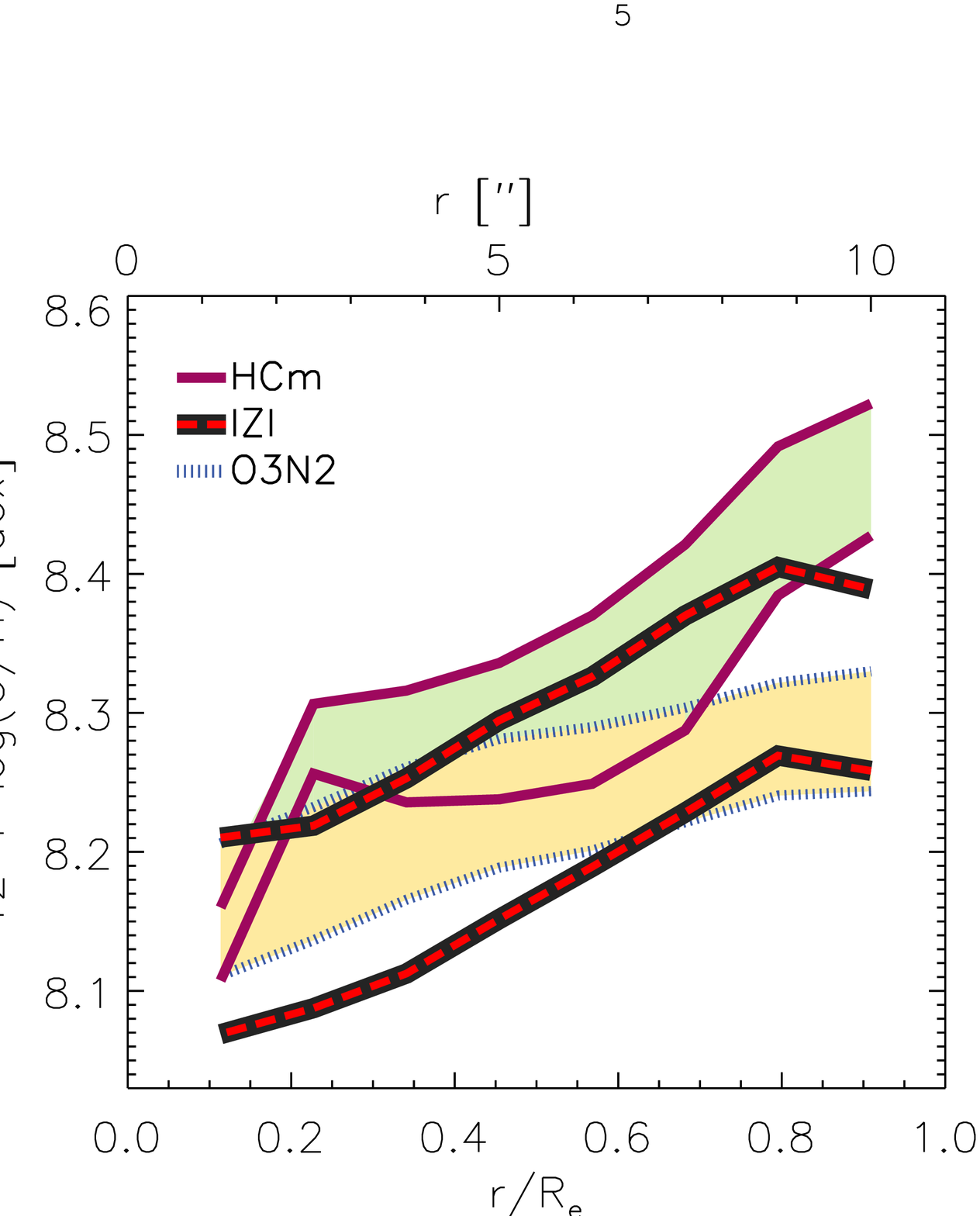}
 \caption{Comparison between the abundance gradients obtained with the three methods: O3N2 (dotted line), \texttt{IZI} 
 (dashed line), and \texttt{HCm} (solid line).}
\label{fig:app:met_grad}
\end{center}
\end{figure}

Lastly, in Fig. \ref{fig:app:met_grad} we plot the oxygen abundance gradients derived with 
the three methods.
\texttt{IZI} and the O3N2 calibrator provide very similar radial variations of O/H, while 
\texttt{HCm} derives a steeper gradient.
Overall the outputs of the different diagnostics are comparable within the uncertainties, 
and they all lead to the same conclusion that we illustrated in Sect. \ref{subsec:metal}.
In the lack of applicability of the direct method, all the diagnostics that we tested are 
consistent with the presence of an inverted abundance gradient in VCC~324.

\bsp	
\label{lastpage}
\end{document}